\documentclass[aps,pra,twocolumn,groupedaddress,showpacs,amsfonts]{revtex4-1}
\usepackage{amsfonts}
\usepackage{amsmath}
\usepackage{amssymb}
\usepackage{graphicx}
\usepackage{epsf}
\usepackage{epsfig}
\usepackage{tabularx}

\newcommand{\be}{\begin{equation}}
\newcommand{\ee}{\end{equation}}
\newcommand{\bem}{\begin{multline}}
\newcommand{\eq}[1]{Eq.~\eqref{#1}}
\newcommand{\seq}[1]{Sec.~\ref{#1}}

\begin{document}


\title{Quasi-bound states in periodically driven scattering}

\author{H. Landa$^1$}
\email{e-mail: haggaila@gmail.com}
\affiliation{$^1$Univ.~Paris Sud, CNRS, LPTMS, UMR 8626, Orsay 91405, France}

\begin{abstract}

We present an approach for obtaining eigenfunctions of periodically driven time-dependent Hamiltonians. Assuming an approximate scale separation between two spatial regions where different potentials dominate, we derive an explicit expansion for scattering problems with mixed cylindrical and spherical symmetry, by matching wavefunctions of a periodic linear drive in the exterior region to solutions of an arbitrary interior potential expanded in spherical waves. Using this method we study quasi-bound states of a square-well potential in three dimensions subject to an axial driving force. In the nonperturbative regime we show how eigenfunctions develop an asymptotic dressing of different partial waves, accompanied by large periodic oscillations in the angular momentum and a nonmonotonous dependence of the decay rate on the drive strength. We extend these results to the strong driving regime near a resonant intersection of the quasi-energy surfaces of two bound states of different symmetry. Our approach can be applied to general quantum scattering problems of particles subject to periodic fields.
\end{abstract}


\maketitle

\section{Introduction}

The main object of this paper is a time-dependent Schr\"{o}dinger equation of the general form
\be i\dot{\phi }\left(\vec{r},t\right)=\left[-\frac{1}{2} \nabla ^{2} +V_{L} \left(\vec{r}\right)+V_{R} \left(\vec{r},t\right)\right]\phi \left(\vec{r},t\right)\label{iphidotLR},\ee
where the potential $V_{R}$ is $\pi$-periodic in time (in units such that the fundamental angular frequency is $2$, and $\hbar=m=1$).

Various Floquet approaches have been developed for studying  problems with similar formal equations in different parameter regimes, and \seq{Sec:FloquetOverview} gives a brief overview of relevant works. Here we present an approach for obtaining approximate eigensolutions of \eq{iphidotLR}, starting with explicitly known solutions for each separate Schr\"{o}dinger equation with one potential $V_{j}$, $j\in \left\{L,R\right\}$. Assuming a scale separation between two regions where either one of the two potentials dominates, quasi-periodic Floquet eigenfunctions of \eq{iphidotLR} are obtained by a time-dependent matching of the wavefunctions, described in \seq{Sec:Matching}.

As the main result of the paper, in \seq{Sec:Matching3D} we present a new explicit expansion for problems with mixed cylindrical and spherical symmetry,
\be i\dot{\phi } \left(\vec{r} ,t\right)=\left[-\frac{1}{2} \nabla ^{2} +V_{{\rm int}}\left(\left|\vec{r} \right|\right)-\ddot{\vec{F}}\left(t\right)\cdot\vec{r}\, \right]\phi \left(\vec{r} ,t\right)\label{iphidot0},\ee
where $V_{{\rm int}}$ is important up to some characteristic distance from the origin, and $\ddot{\vec{F}}\left(t\right)$ is the periodic driving force. Relevant examples for such a geometry include, as detailed below, an atom in a linearly polarized laser field, a quantum particle interacting with a periodically-driven semiclassical scatterer (e.g. a cold atom interacting with an ion in a Paul trap), and interacting cold atoms or molecules that are subject to oscillating fields. The generalization to other settings, e.g. with a time-dependent potential in the interior region, or a potential with spherical symmetry in the exterior region, is straightforward.
The presented matching conditions give immediately the complete spatial information on the wavefunctions. The explicit usage of analytic wavefunctions in each region gives access to fine details of the spectrum which may be hard to locate otherwise, and in particular the widely used Quantum Defect Theory (QDT) \cite{PhysRevA.26.2441,Seaton1983} can naturally be used in the interior region. Our approach is nonperturbative in both potentials $V_{j}$, but neglects the effect of either potential in some region of space. Therefore the obtained solutions can be considered, if necessary, as a starting point for an expansion that will treat the neglected contributions.

Finally, in \seq{Sec:Numerics} we employ our expansion to calculate solutions of \eq{iphidot0} with a spherical square-well potential, and demonstrate general phenomena in the nonperturbative regime, e.g. nonmonotonous parametric dependence of the decay rate out of the well, large periodic oscillations of observables, and the resonant intersection of the quasi-energy surfaces of two bound states of different symmetry. 

\section{Overview of Floquet scattering}\label{Sec:FloquetOverview}

If we consider $-\ddot{\vec{F}}\left(t\right)$ to be a monochromatic electric field amplitude, and $V_{{\rm int}}\left(\left|\vec{r} \right|\right)$ as the Coulomb potential for an electron, \eq{iphidot0} describes an atom in an AC field (the AC-Stark effect), written in the length-gauge within the dipole approximation. In \cite{Yajima1982,Yajima1983} it is proved that with \eq{iphidot0}, the bound states of $V_{{\rm int}}$ (under general assumptions) become resonances with an imaginary part which depends as a power-law on the amplitude of the perturbation -- indeed the proof is perturbative in the electric field amplitude.

This result is in fact general -- for a Hamiltonian with a continuous spectrum of scattering states, the bound states will generally turn into resonances under the effect of a periodic perturbation \cite{Yafaev1991}. The reason is that the periodic perturbation makes every bound state with energy $\left(-\left|\omega\right|\right)$ resonant with unbound states from the continuum of positive energy states, under absorption of at least $n$ quanta from the perturbing potential (whose frequency is $2$), where \be\left(-\left|\omega\right|\right)+2n >0\label{epsilon2n},\ee
and $2n$ gives the exponent of the power-law dependence of the resonance width on the perturbation amplitude.

To be contrasted with the above picture, almost all of the bound states of a time-independent Hamiltonian can strictly survive the addition of a periodic perturbation, if the unperturbed Hamiltonian has a discrete spectrum (of isolated eigenvalues of finite multiplicity \cite{casati1989quantum}), a pure point spectrum (of discrete eigenvalues \cite{Howland1989,Howland1989b,Combescure1990}), or a bounded continuous spectrum (in which case the perturbation must obey certain conditions \cite{Martin1998}). The discreteness or boundedness of the spectrum in these cases stabilizes the spectrum (typically except under some specific resonances with the external field), for any strength of the perturbation, and in these cases the proofs are nonperturbative.

Nonperturbative studies of eqs.~\eqref{iphidotLR}-\eqref{iphidot0} have received a lot of attention within the intense-laser literature \cite{0034-4885-60-4-001}, where \eq{iphidot0} is designated as being in the Kramers-Hanneberger (KH) frame. A variety of approaches have been developed for tackling this problem, focusing on different physical questions and in different parameter regimes of laser frequency, intensity and polarization. In this language, \eq{epsilon2n} expresses the fact that the rate of $n$-photon ionization is proportional to the $n$-th power of the field intensity, a result derived already in the early days of the field within Keldysh theory \cite{Keldysh1965}.

A very general Green's function approach was developed in \cite{PhysRevLett.52.613,PhysRevA.37.4536} and solved for Coulomb scattering of electrons, by neglecting all time-dependent terms except the leading-order averaged term. Known as the KH approximation, this approach is suitable in the regime where the frequency and intensity of the oscillating field are much higher than the atomic potential (in atomic units). In this limit, the perturbative picture of ionization rate which increases with intensity breaks, and the significant distortion of the effective (``dressed'') potential seen by the electron leads to the remarkable phenomenon of stabilization of the atom against ionization. Ref. \cite{gavrila2002atomic} gives a comprehensive review of the works related to this effect. The interest in the ``KH atom'' has been renewed in recent years following experimental results \cite{eichmann2009acceleration,eichmann2013observing,0953-4075-47-20-204014} and theoretical investigations \cite{PhysRevLett.110.253001,PhysRevA.90.023401}. Recent works have also revisited the systematic expansion of an effective time-independent Hamiltonian in the high-frequency limit \cite{PhysRevLett.91.110404,PhysRevA.68.013820}, and the effects related to the potential's initial phase \cite{PhysRevA.76.013421}.

For lower fields and frequencies, a wealth of techniques have been applied in the field. Photon absorption or emission processes which couple different scattering channels were treated by numerically integrating the close-coupled equations \cite{PhysRevLett.59.872,PhysRevA.44.R5343}, and employing a QDT approach to extrapolate scattering cross sections \cite{PhysRevA.36.5178,PhysRevA.43.1512}. Various effects related to this rich problem have been analyzed in simpler settings \cite{PhysRevA.37.98,PhysRevA.40.5614,PhysRevA.45.6735,moiseyev1991multiphoton,ben1993creation,timberlake2001phase,emmanouilidou2002floquet,PhysRevA.69.062105,PhysRevA.71.012102,PhysRevA.85.023407}, including the appearance and annihilation of bound states in the dressed potential, avoided crossing of resonances and their behaviour in phase-space, resonant coupling between internal levels, and nonmonotonic ionization rates (as laser intensity is increased). 
Within R-matrix theory in the Floquet setting, space is divided into two regions and the solutions (typically obtained numerically within each region) are connected at the boundary \cite{burke1991r}, and new extensions of this approach have been recently suggested \cite{PhysRevA.86.043408}. Numerical integration techniques have evolved in complexity and sophistication \cite{0034-4885-60-4-001} and continue to be improved, and in particular, there is interest in calculating and directly probing the angular distribution of photoelectron spectra \cite{PhysRevA.78.043403,Morales11102011}.
AC Stark shifts of trapped atoms have recently been modeled and measured in \cite{markert2010ac}.

In a more general setting, the formalism for treating Hamiltonians periodic in time, using Floquet theory, is well known
\cite{PhysRev.138.B979,PhysRevA.7.2203,tannor2007introduction}. 
The periodicity of the Hamiltonian allows defining an extended Hilbert space in position and time, in which the scalar product is defined to include integration over the temporal period. This extended Hilbert space can be spanned by set of spatially orthogonal wavefunctions and a Fourier basis for time-periodic functions, e.g. by all wavefunctions of a specific $\pi$-periodic Hamiltonian. Then, any other $\pi$-periodic Hamiltonian can be expanded using such a basis for the extended Hilbert space, and all of the tools of time-independent quantum theory are available, which can be powerful in many scenarios, e.g. for employing perturbation theory. 
Time-dependent perturbation theory is also widely used \cite{RevModPhys.44.602}, and usually the interest is in transition rates between asymptotically time-independent states.

To conclude this section, we briefly mention the recent interest in atomic systems in the ultracold regime (i.e.~with temperature $\lesssim 1\,{\rm mK}$, \cite{carr2009cold}), which are trapped by oscillating fields. Overlapping a trap for neutral atoms with a periodically driven Paul trap for ions \cite{WinelandReview}, was suggested in \cite{PhysRevA.67.042705} and realized first in \cite{PhysRevLett.102.223201}, followed by the demonstration of a trapped ion immersed in a dilute atomic Bose-Einstein condensate \cite{zipkes2010trapped,PhysRevLett.105.133202}, and many other experiments. The effect of the periodic drive of the ion has been analyzed for classical collisions with the atom \cite{PhysRevLett.109.253201}, for quantum scattering employing a master equation description \cite{PhysRevA.91.023430}, and for an ion and atom in separate traps \cite{PhysRevA.85.052718}.
As mentioned before, Quantum Defect Theory (QDT) constitutes one of the most common theoretical tool for modelling the short-range part of the interaction in atomic scatterring setups, and continues to evolve \cite{PhysRevA.78.012702,PhysRevA.79.010702,PhysRevA.75.053601,PhysRevA.80.012702,PhysRevLett.104.213201,PhysRevA.84.042703,idziaszek2011multichannel,PhysRevA.88.022701,PhysRevLett.110.213202,PhysRevA.87.032706}, together with new models and methods \cite{PhysRevA.73.063619,PhysRevA.82.042712}, applied to many-body states as well \cite{PhysRevA.90.033601,schurer2015capture}. As a last example for a driven system with two-body interaction we mention  polar molecules in AC traps \cite{PhysRevLett.94.083001,tokunaga2011prospects}.

\section{Quasi-bound states in time-periodic potentials}

In this section we formulate a method for finding wavefunctions of \eq{iphidotLR}, with a time-independent potential $V_{L} \left(\vec{r}\right)$ which is assumed to be significant inside some interior region $\left|\vec{r}\right|<d$, and a time-dependent $\pi$-periodic potential $V_{R} \left(\vec{r},t\right)$ which dominates in the exterior region $\left|\vec{r}\right|>d$. The essential assumptions at the basis of the presented approach are that the wavefunctions of each of the potentials can be found explicitly, and that there is some meaning to dividing space into the interior and exterior regions, even if only as a (zeroth-order) approximation. We focus in this paper on finding quasi-bound states (resonances), while scattering states follow the same expansion with just a redefinition of the unknown parameters.

\subsection{Matching conditions for Floquet-expanded wavefunctions}\label{Sec:Matching}

To simplify the basic expressions, we take a one-dimensional (1D) notation for the derivation in this subsection, starting with the 1D equation corresponding to \eq{iphidotLR},
\be i \dot{\phi }\left(x,t\right)=\left[-\frac{1 }{2} \nabla ^{2} +V_{L} \left(x\right)+V_{R} \left(x,t\right)\right]\phi \left(x,t\right). \ee
To formulate the matching conditions of the quasi-periodic wavefunction $\phi \left(x,t\right)$ at a boundary point $d$ which separates the regions of the left potential $V_{L} \left(x\right)$ and the right potential $V_{R} \left(x,t\right)$, we consider the ansatz
\be \phi \left(x,t\right)=\left\{\begin{array}{cc} {\sum_{n}a_{2n} \phi _{L,\omega +2n} \left(x\right)e^{-i\left(\omega +2n\right)t}  } & {x<d} \\ \\{\sum_{n}b_{2n} \phi _{R,\omega +2n}^{\pi } \left(x,t\right)e^{-i\left(\omega +2n\right)t}  } & {x>d} \end{array}\right.\label{phiansatz}\ee
where $\phi _{L,\omega +2n} \left(x\right) $ and $\phi _{R,\omega +2n}^{\pi } \left(x,t\right) $ (the latter being $\pi$-periodic) are solutions of the Schr\"{o}dinger equation with potential $V_L\left(x\right)$ and $V_R\left(x,t\right)$ respectively, and energy (quasi-energy) $\omega+2n$. For notational simplicity, the above summation does not indicate explicitly a summation over any degeneracy of the wavefunctions, which must involve independent matching coefficients. In the 1D case this can include left- and right-going waves (if the boundary conditions allow), and in higher dimensions there could be summation over other quantum numbers. It is also assumed here that some prescribed boundary conditions at $x\to \pm \infty $ are already included in $\phi _{L,\omega +2n} $ and $\phi _{R,\omega +2n}^{\pi }$. We omit the explicit range of Fourier summations on integers $n\in \mathbb{Z}$. The wavefunction $\phi \left(x,t\right)$ is parametrized by $\omega $ which can be chosen in the range $-2<\omega \le 0$, however $\omega $ does not determine uniquely the wavefunction -- there can be different functions with the same value of $\omega $ (but different coefficients).

 The matching conditions at $d$ are
\be \phi _{L} \left(d\right)=\phi _{R} \left(d,t\right),\qquad \nabla \phi _{L} \left(d\right)=\nabla \phi _{R} \left(d,t\right)\label{matchingconditions},\ee
and the normalization applicable to a square-integrable wavefunction is
\be \int \phi ^{*} \left(x,t\right)\phi \left(x,t\right)dx =1.\ee
Expanding the functions at the matching point we write,
\be \phi _{L,\omega +2n} \left(d\right)= c_{2n},\ee
\be \phi _{R,\omega +2n}^{\pi } \left(d,t\right)=\sum _{k}d_{2k} e^{-i2kt},\ee
so the first matching condition of \eq{matchingconditions}, implies
\be \sum _{n}a_{2n} c_{2n} e^{-i2nt} = \sum _{j,k }d_{2k}  b_{2j} e^{-i2\left(j+k\right)t},\ee
which gives
\be c_{2n} a_{2n} =\sum _{j }d_{2\left(n-j\right)} b_{2j}.\ee
A similar expansion for the gradients,
\be \nabla \phi _{L,\omega +2n}^{\pi } \left(d\right)= f_{2n},\ee
\be \nabla \phi _{R,\omega +2n}^{\pi } \left(d,t\right)=\sum _{k }g_{2k} e^{-i2kt},\ee
gives
\be f_{2n} a_{2n} =\sum _{j }g_{2\left(n-j\right)} b_{2j}.\ee

The two matching relations can be written in matrix form (once a finite truncation has been applied),
\be C\vec{a}=D\vec{b},\qquad F\vec{a}=G\vec{b}.\ee
where $\vec{a},\vec{b}$ denote the expansion coefficients in vector notation and $C,D,F,G$ are matrices. By writing the two equations in block form
\be K\left(\begin{array}{c} {\vec{a}} \\ {\vec{b}} \end{array}\right)\equiv \left(\begin{array}{cc} {C} & {-D} \\ {F} & {-G} \end{array}\right)\left(\begin{array}{c} {\vec{a}} \\ {\vec{b}} \end{array}\right)=0,\ee
the compatibility of the two matching conditions implies the vanishing of (at least one) eigenvalue (or, more generally, singular value in the SVD decomposition) of $K\left(\omega \right)$, with the corresponding kernel vector then giving the expansion coefficients. The same arguments can be applied to the smaller matrix (since $C$ and $F$ are assumed diagonal in the current expansion, and would in general be invertible)
\be FC^{-1} D\vec{b}=G\vec{b}\qquad \Rightarrow \left(G-FC^{-1} D\right)\vec{b}=0,\ee
whose kernel vectors give the exterior region coefficients $\vec{b}$, from which $\vec{a}$ immediately follows.

Defining 
\be P_{2k,2n} =\int _{d}^{\infty }dx\phi _{R,\omega +2k}^{\pi } \left(x,0\right)^{*} \phi _{R,\omega +2n}^{\pi } \left(x,0\right)\ee
and similarly,
\be Q_{2k,2n} =\int _{-\infty }^{d}dx\phi _{L,\omega +2k}^{\pi } \left(x,0\right)^{*} \phi _{L,\omega +2n}^{\pi } \left(x,0\right),\ee
we get for the normalization condition (which can be evaluated at $t=0$), the bilinear expression 
\bem {\int _{-\infty }^{\infty }\phi ^{*} \phi  dx=\int _{-\infty }^{d}\phi ^{*} \phi  dx+\int _{d}^{\infty }\phi ^{*} \phi  dx} \\=\sum _{k,n}a_{2k} ^{*} Q_{2k,2n} a_{2n}   +\sum _{k,n}b_{2k} ^{*} P_{2k,2n} b_{2n}   =1,\label{phinormalization}\end{multline}
or in matrix form,
\be \vec{a}^{\dag } Q\vec{a}+\vec{b}^{\dag } P\vec{b}=1,\ee
so that normalization can be guaranteed by dividing $\vec{a},\vec{b}$ by the square-root of the l.h.s. We note that this normalization is relevant only if the entire wavefunction is square-integrable, and we will discuss the case of wavefunctions with free-particle components in the next section [following \eq{normalization3D}].

\subsection {3D matching of wavefunctions with mixed cylindrical and spherical symmetry}\label{Sec:Matching3D}

In this subsection we write the matching conditions for the 3D problem
\be i \dot{\phi }\left(\vec{r},t\right)=\left[-\frac{1 }{2} \nabla ^{2} +V_{{\rm int}}\left(\left|\vec{r} \right|\right)-\ddot{\vec{F}}\left(t\right)\cdot \vec{r} +V_{1} \left(t\right)\right]\phi \left(\vec{r},t\right)\label{iphidot3D}\ee
with a spherically-symmetric potential in the interior region, and a linear (periodic) drive in the exterior region, where $V_{1} \left(t\right)$ is added for convenience as detailed below.

 For the general linearly-driven time-dependent Schr\"{o}dinger equation (in the exterior region)
\be i \dot{\phi }\left(\vec{r},t\right)=\left[-\frac{1 }{2} \nabla ^{2} -\ddot{\vec{F}}\left(t\right)\cdot \vec{r} +V_{1} \left(t\right)\right]\phi \left(\vec{r},t\right)\label{ihbarphidot3Dext},\ee
 a family of solutions can be written in the form
\be \phi \left(\vec{r},t\right)\propto e^{ i\vec{q}\left(t\right)\cdot \vec{r}-ig\left(t\right)},\ee
with
\be \vec{q}\left(t\right)=\dot{\vec{F}} +\vec{k}\ee
where $\vec{k}$ is the (possibly complex) constant of integration, and 
\be {g\left(t\right)=\frac{1 }{2} \vec{k}^{2} t}\\+\int _{}^{t}\left[ \vec{k}\cdot \dot{\vec{F}}\left(t'\right)+\frac{1}{2}\dot{\vec{F}}\left(t'\right)^{2} +V_{1} \left(t'\right) \right]dt'.\label{gtint}\ee

Below we will need the cylindrical waves, i.e. the solutions of the time-independent free particle Hamiltonian in cylindrical coordinates $\left(\rho,z,\varphi\right)$, defined by
\be \chi _{m}^{\left(1,2,J\right)} \left(\vec{r};k,\alpha \right)= e^{im\varphi }H_{m}^{\left(1,2,J\right)} \left(k\rho \sin \alpha \right)e^{ikz\cos \alpha },\label{chim12J}\ee
where $m$ is the magnetic quantum number, $k$ is the wavenumber which can in general be complex, and $\alpha$ is a complex parameter. $H_{m}^{\left(1,2,J\right)}$ is a Hankel function of the first or second kind (corresponding to outgoing and incoming traveling waves respectively), or a Bessel function (which we denote with a superscript $J$).

Specializing to the case that the drive is $\pi$-periodic and coaxial at any time, we can choose a fixed cylindrical coordinate system in which 
\be {\vec{F}}\left(t\right)={F}^{\pi } \left(t\right)\hat{z}\label{Fpi}.\ee
We will further simplify the current expressions by taking $V_1\left(t\right)$ of eqs.~\eqref{iphidot3D}-\eqref{ihbarphidot3Dext} to cancel the $\vec{k}$-independent term in \eq{gtint}, so that outgoing and incoming traveling-wave solutions to \eq{ihbarphidot3Dext} can be written using \eq{chim12J} in the form
\bem \phi _{R,m}^{\left(1,2\right)} \left(\vec{r},t;k,\alpha \right)\\ \propto e^{-i\frac{1}{2} k^{2} t} e^{i\dot{F}^{\pi } \left(t\right)z} \chi _{m}^{\left(1,2\right)} \left(\vec{r};k,\alpha \right)e^{-iF^{\pi } \left(t\right)k\cos \alpha }\label{phiRm12}.\end{multline}


Fixing the magnetic quantum number $m$ which is conserved, and defining $k_{2j}=\sqrt{2\left(\omega+2j\right)}$, the most general wavefunction that solves \eq{ihbarphidot3Dext} for the exterior region $\left|\vec{r}\right|>d$ is
\be \phi _{R,m} \left(\vec{r},t\right)=\sum _{\substack{j\\a=1,2}}\int_{C_a} d\alpha \sin\alpha b_{2j}^{\left(a\right)} \left(\alpha \right)\phi _{R,m}^{\left(a\right)} \left(\vec{r},t;k_{2j} ,\alpha \right),\label{phimexterior}\ee
which takes at each value of $k_{2j}$ a superposition of outgoing and incoming cylindrical waves, parameterized by integrals in complex $\alpha$-plane along two contours $C_{a}$ with weight functions $b_{{2j}}^{\left(a\right)}\left(\alpha\right)$, both to be determined in the following.

The above expansion becomes useful by using a representation of the spherical Hankel function of the first kind as an integral over cylindrical waves \cite{DanosMaximon,bostrom1991transformation} in the form
\bem h_{l}^{\left(1\right)} \left(kr\right)P_{l}^{m} \left(\cos \theta \right)e^{im\varphi } \\=\int _{C_1}d\alpha \sin \alpha \frac{1}{2} \left(-i\right)^{l-m} P_{l}^{m} \left(\cos \alpha \right)\chi _{m}^{\left(1\right)} \left(\vec{r};k,\alpha \right)\label{hlexpansion}\end{multline}
where $P_l^m$ are the associated Legendre polynomials and the directed contour of integration $C_1$ depends on $k$. For $k$ with a positive imaginary part we must take $C_1=\pi /2+i\left(\infty,-\infty\right)$, and $h_{l}^{\left(1\right)} \left(kr\right)$ then decays asymptotically as $e^{-\left|k\right|r}/r$ (we note that $\chi _{m}^{\left(1\right)} \left(\vec{r};k,\alpha \right)$ diverges at $\rho \to 0$, however the integral, which gives $h_{l}^{\left(1\right)} \left(kr\right)$, is well defined for any $\rho >0$, and decays for $r\to \infty $, which is just what we need). For $k$  real and positive the contour of integration is given by $C_1=i\left(\infty ,0\right)+\left[0,\pi \right]+\left\{\pi +i\left(0,-\infty \right)\right\}$. 

This directs us to take the arbitrary weight function for outgoing waves in \eq{phimexterior} to be of the form 
\be b_{{2j} }^{\left(1\right)} \left(\alpha \right)=-\sum _{l_{1} }b_{2j,l_{1} } S_{R,2j,l1}^{\left(1\right)} P_{l_{1} } \left(\cos \alpha \right),\label{be2jm} \ee
 with $b_{2j,l_{1} }$ to become matching coefficients and the constants $S_{R,2j,l1}^{\left(1\right)}$ are determined by boundary conditions at infinity as detailed below. In App.~\ref{Sec:Derivations1} we show that each term in the summation of \eq{be2jm}, when plugged into the integral over $C_1$ [in \eq{phimexterior}], can be written in the following form;
\bem {\int_{C_1} d\alpha \sin\alpha b_{{2j} }^{\left(1\right)} \left(\alpha \right)\phi _{R,m}^{\left(1\right)} \left(\vec{r},t;k_{2j} ,\alpha \right)}=\\ {-e^{-i\frac{1}{2} k_{2j} ^{2} t}\sum _{l_{1},l }b_{2j,l_{1} }S_{R,2j,l1}^{\left(1\right)} R_{2j,l,l_{1} } ^{\left(1\right)}\left(r,t\right) Y_{l}^{m} \left(\theta ,\varphi \right)  }\label{intphiRexpansion} \end{multline}
where $Y_{l}^m$ are spherical harmonics, and the outgoing and incoming radial functions in the exterior region are defined by
\bem R_{2j,l_{1} ,l}^{\left(a\right)} \left(r,t\right)=\\\sum _{l_{2} ,l_{3} ,l_{4} }c_{l_{1} ,l_{2} ,l_{3} ,l_{4} ,l} j_{l_{2} } \left(F^{\pi } \left(t\right)k_{2j} \right)j_{l_{4} } \left(\dot{F}^{\pi } \left(t\right)r\right) h_{l_{3} }^{\left(a\right)} \left(k_{2j} r\right)\label{R2jl1la}\end{multline}
with the coefficients $c_{l_{1} ,l_{2} ,l_{3} ,l_{4} ,l} $ being defined in \eq{cl1l2l3l4l}.

For any value of $k$, we have similarly to \eq{hlexpansion}
\bem j_{l} \left(kr\right)P_{l}^{m} \left(\cos \theta \right)e^{im\varphi } =\\ \int _{\left[0,\pi \right]}d\alpha \sin \alpha \frac{1}{2} \left(-i\right)^{l-m} P_{l}^{m} \left(\cos \alpha \right)\chi _{m}^{\left(J\right)} \left(\vec{r};k,\alpha \right),\label{jlexpansion}\end{multline}
where $j_l$ is a spherical Bessel function, and \eq{jlexpansion} allows to express the spherical Hankel function of the second kind (for $k$ with nonnegative imaginary part) by using $h_{l}^{\left(2\right)} =2j_{l} -h_{l}^{\left(1\right)}$. Therefore, using the fact that the $l,m$-dependent coefficients in \eq{hlexpansion} and \eq{jlexpansion} are identical, we can replace in \eq{phimexterior} the integral over the contour $C_2$ (which need not be further specified) by an expression identical in form to the expansion in \eq{intphiRexpansion}, with the outgoing waves replaced by (minus) incoming waves. The wavefunction expansion in 3D analogous to \eq{phiansatz} is then
\begin{widetext}
\be \phi _{m} \left(\vec{r},t\right)=\left\{\begin{array}{ccc} {\sum_{n,l}a_{2n,l} e^{-i\left(\omega +2n\right)t} \phi _{L,\omega +2n,l}\left(r\right) Y_{l}^m\left(\theta,\varphi\right)  } & & {\left|\vec{r}\right|<d} \\\\ {\sum _{ j,l_{1}} b_{2j,l_{1} }e^{-i\left(\omega +2j\right)t}\sum _{l}\phi_{R,2j,l_1,l}^{\pi} \left(r,t\right)Y_{l}^m \left(\theta ,\varphi \right)} & & {\left|\vec{r}\right|>d} \end{array}\right.\label{phi3Dansatz},\ee
with
\be \phi_{R,2j,l_1,l}^{\pi} \left(r,t\right)= S_{R,2j,l_1}^{\left(2\right)} R_{2j,l_{1},l }^{\left(2\right)} \left(r,t\right)-S_{R,2j,l_1}^{\left(1\right)}R_{2j,l_{1},l }^{\left(1\right)} \left(r,t\right).  \ee
In the interior region the wavefunction $\phi _{L,\omega +2n,l}\left(r\right)$ is a solution with energy $\omega+2n$ of the Schr\"{o}dinger equation with potential $V_L\left(\vec{r}\right)$, expressed in spherical coordinates $\left(r,\theta,\varphi\right)$, and the summation includes all partial waves and energies. In the exterior region we have incorporated the effect of boundary conditions at infinity into $S_{R,2j,l_1}^{\left(a\right)}$, the choice of which will be detailed at the end of the current section.

Therefore, equating the wavefunction on the surface of the sphere $\left|\vec{r}\right|=d$ we have
\bem \sum _{n,l}e^{-i\left(\omega +2n\right)t} a_{2n,l}  \phi _{L,\omega +2n,l} \left(d\right) Y_{l}^m  =\sum _{ j,l_{1}} e^{-i\left(\omega +2j\right)t}b_{2j,l_{1} }\sum _{l} \phi_{R,2j,l_1,l}^{\pi} \left(d,t\right) Y_{l}^m \\  =\sum _{ j,l_{1}} e^{-i\left(\omega +2j\right)t}b_{2j,l_{1} }\sum _{l,p} d_{2j,l_{1},l,2p } e^{-i2pt}Y_{l}^m     =\sum _{ n,l} e^{-i\left(\omega +2n\right)t}\sum _{j,l_1}b_{2j,l_{1} } d_{2j,l_{1},l,2\left(n-j\right) } Y_{l}^m   \end{multline}
where $d_{2j,l_{1},l,2p } $ are the expansion coefficients of the Fourier series of $\phi_{R,2j,l_1,l}^{\pi} \left(d,t\right)$, which in general must be obtained numerically.
Then the first matching condition is
\be \phi _{L,\omega +2n,l} \left(d\right) a_{2n,l} =\sum _{j,l_1} d_{2j,l_{1},l,2\left(n-j\right) } b_{2j,l_{1} }.\ee
Since we equate the interior and exterior wavefunctions on the surface of a sphere, we need to consider only the radial derivatives. This gives
\be \partial _{r} \phi _{L,\omega +2n,l} \left(d\right)a_{2n,l} =\sum _{j,l_1} g_{2j,l_{1},l,2\left(n-j\right) } b_{2j,l_{1} },\ee
with $g_{2j,l_{1},l ,2p}  $ the expansion coefficients of the Fourier series of $\partial _{r} \phi_{R,2j,l_1,l}^{\pi} \left(d,t\right)$. The latter derivative can be written explicitly as
\be \begin{array}{l} {\partial _{r} R_{2j,l_{1} ,l }^{\left(a\right)} \left(r,t\right)=\sum _{l_{2} ,l_{3} ,l_{4} }c_{l_{1} ,l_{2} ,l_{3} ,l_{4} ,l} j_{l_{2} } \left(F^{\pi }\left(t\right) k_{2j} \right)} \\ \\ {\qquad  \times\left\{\dot{F}^{\pi }\left(t\right) j_{l_{4} +1} \left(\dot{F}^{\pi }\left(t\right) r\right)h_{l_{3} }^{\left(a\right)} \left(k_{2j} r\right)+j_{l_{4} } \left(\dot{F}^{\pi }\left(t\right) r\right)\left[\frac{1}{r}\left(l_{3} +l_{4} \right) h_{l_{3} }^{\left(a\right)} \left(k_{2j} r\right)-k_{2j} h_{l_{3} +1}^{\left(a\right)} \left(k_{2j} r\right)\right]\right\}.} \end{array}\ee
For the normalization integrals performed at $t=0$ as in \eq{phinormalization}, we find using the orthogonality of $Y_{l}^m$,
\be 1= \sum _{n,n'}a_{2n',l}^* a_{2n,l}\int_0^d dr r^2 \phi_{L,\omega+2n',l}^*  \phi_{L,\omega+2n,l} +\sum _{j,j',l_{1} ,l_{1} ^{{'} } }b_{2j',l_{1} ^{{'} } } ^{*} b_{2j,l_{1} } \sum _{l} \int_d^\infty dr r^{2}\left[\phi_{R,2j',l_{1} ^{{'} } ,l} ^{\pi} \right]^*\phi_{R,2j,l_{1} ,l }. \label{normalization3D}\ee
\end{widetext}

In App.~\ref{Sec:Derivations3} we lay down for completeness the expansion of integrals which are required in order to calculate expectation values of general rank-0, -1, and -2 tensor operators (we restrict the expressions to axially symmetric wavefunction with $m=0$). The expectation value of any time-independent (or $\pi$-periodic) operator is always $\pi$-periodic for the Floquet eigenstates. The normalization integrals in \eq{normalization3D} are a special case of \eq{I1integral}, which can be evaluated at $t=0$ if the wavefunction is square-integrable. However, for expansions which contain free-particle components (discussed in the following), when we can only integrate over the bound components of the wavefunction, the normalization integral is $\pi$-periodic because the relative weight of the nonnormalizable components oscillates in time. In this case one must divide expectation value integrals by the squared norm, both of which being $\pi$-periodic functions.

In the interior region, the values of the above integrals can be obtained without explicitly performing the integration, directly from the wavefunctions and their gradients at the matching point. This can useful especially when the interior wavefunctions are not explicitly known close to the origin, but rather are determined within a QDT formulation \cite{Seaton1983,Gao2008}.
We will use the following general notation for wavefunctions in spherical coordinates;
\be \phi\left(\vec{r}\right) = \frac{1}{r}u\left({r}\right)Y_{l}^{m} \left(\theta ,\varphi \right),\label{phi1ruY}\ee
such that $u\left({r}\right)$ is a radial function obeying a one-dimensional Schr\"{o}dinger equation with an effective potential which includes the centrifugal barrier.
The projection of two eigenfunctions $\phi_1$ and $\phi_2$ of the interior Hamiltonian with energies $\varepsilon_1$ and $\varepsilon_2$ correspondingly, is shown in App.~\ref{Sec:Derivations2} to be given by
\be 2\,\mathfrak{Re}\int _{0}^{d }\phi _{1} ^{*} \phi _{2} r ^{2} dr = \left(\varepsilon _{1} -\varepsilon _{2} \right)^{-1} \, \mathfrak{Re}\left. \left\{u_{1} ^{*} u_{2} ^{{'} } -u_{2} ^{*} u_{1} ^{{'} } \right\}\right|_{d },\label{phi1phi2int}\ee
where $u_1^{'}\equiv \partial _{r} u_{1}$, and  $\varepsilon _{1}$, $\varepsilon _{2}$ are assumed to have equal imaginary parts. The left-hand side of \eq{phi1phi2int} gives the integrals required for the normalization, with the factor of 2 relevant for the off-diagonal projections (when $\phi_1\neq \phi_2$). In the limit of $\phi_1\to \phi_2$ we have
\be \int _{0}^{d }\left|\phi _{1} \right|^{2} r ^{2} dr  = \frac{1}{2}\mathop{\lim }\limits_{\varepsilon _{1} \to \varepsilon _{2} } \left(\varepsilon _{1} -\varepsilon _{2} \right)^{-1} \left. \left[u_{1} ^{*} u_{2} ^{{'} } -u_{2} ^{*} u_{1} ^{{'} } \right]\right|_{d},\ee
which gives the diagonal normalization terms.

Finally, we comment on the constants $S_{R,2j,l_1}^{\left(a\right)}$ in \eq{phi3Dansatz}, which should be chosen in the following way, consistent with the domain of validity of \eq{hlexpansion}. For the partial waves with $\mathfrak{Re}\left\{\omega\right\}+2j<0$ the wavevector $k_{2j}$ can be chosen with a positive imaginary part and setting $S_{R,2j,l_1}^{\left(2\right)}=0$, $ S_{R,2j,l_1}^{\left(1\right)}=1$ gives exponentially decaying outgoing waves. In the terms with $\mathfrak{Re}\left\{\omega\right\}+2j>0$ two types of boundary conditions at infinity can be imposed. If we assume that positive energy states are free (and form a continuum), we should set $S_{R,2j,l_1}^{\left(2\right)}=1$, $S_{R,2j,l_1}^{\left(1\right)}=0$ and search accordingly for resonances with $\mathfrak{Im}\left\{\omega\right\}>0$ and $\mathfrak{Im}\left\{k\right\}>0$, which correspond to incoming waves whose amplitude diverges at infinity. The imaginary part of $\omega$ gives the rate of formation of the resulting quasi-bound state. Otherwise, if there is some potential at infinity which reflects waves inwards, solutions will have $\omega$ purely real, we can set $S_{R,2j,l_1}^{\left(1\right)}=1$, and $S_{R,2j,l_1}^{\left(2\right)}$ gives the relative phase of waves reflecting from the boundary, assuming that it depends only on the energy and the partial-wave angular momentum quantum number $l_1$ of the nondriven problem.

\section{Linear drive with a square-well interaction}\label{Sec:Numerics}

In this section we employ the methods presented in the previous sections to study a model system consisting of a spherically-symmetric square-well potential and a time-dependent periodic linear drive which acts outside of the well. We demonstrate the analysis of general phenomena in the nonperturbative regime.
It is interesting to note that quantum wires and dots \cite{leyronas2001quantum} have been modeled by similar finite-barrier potentials, and the expansion presented here can be used to solve a mixed-type system.

Using the frequency of the periodic drive, $\Omega$, we can define the length and energy scales
\be d_{o} =\sqrt{2\hbar /m \Omega }, \qquad E_o = \hbar\Omega/2, \label{units}\ee
and the variables become nondimensional by rescaling according to 
\be \vec{r}\to \vec{r}/d_{o}, \qquad \vec{k}\to \vec{k}d_{o}, \qquad t\to t\Omega /2,\ee
after which we have explicitly $\hbar=m=1$ and the drive's frequency in these units is $\Omega=2$. With a spherical square-well potential,
\be V_{{\rm well}} \left(\left|\vec{r} \right|\right)=\left\{\begin{array}{cc} {-V_{0} ,} & {\left|\vec{r} \right|<d} \\ {0} & {\left|\vec{r} \right|>d} \end{array}\right.\label{squarewell}\ee
(where $d$ and $V_0$ are nondimensional, measured in the units of \eq{units}), the Schr\"{o}dinger equation in the interior region becomes
\be i\dot{\phi }=\left[-\frac{1}{2} \nabla^{2} +V_{{\rm well}} \left(\left|\vec{r}\right|\right) \right]\phi,\ee
and the regular solution inside the well is a spherical Bessel function,
\be \phi _{\left\{k,l,m\right\}} \left(\vec{r},t\right)\propto e^{-i\left(\frac{1}{2} k^{2} -{V_{0} } \right)t}j_{l} \left(kr\right)Y_{l}^{m}.\ee
In those units, we take the periodic force of \eq{Fpi} to be a simple harmonic drive with amplitude $F_2$,
\be F^{\pi}\left(t\right) = F_2\cos2t.\ee

\begin{figure}[ht]
\center {\includegraphics[width=3.0in]{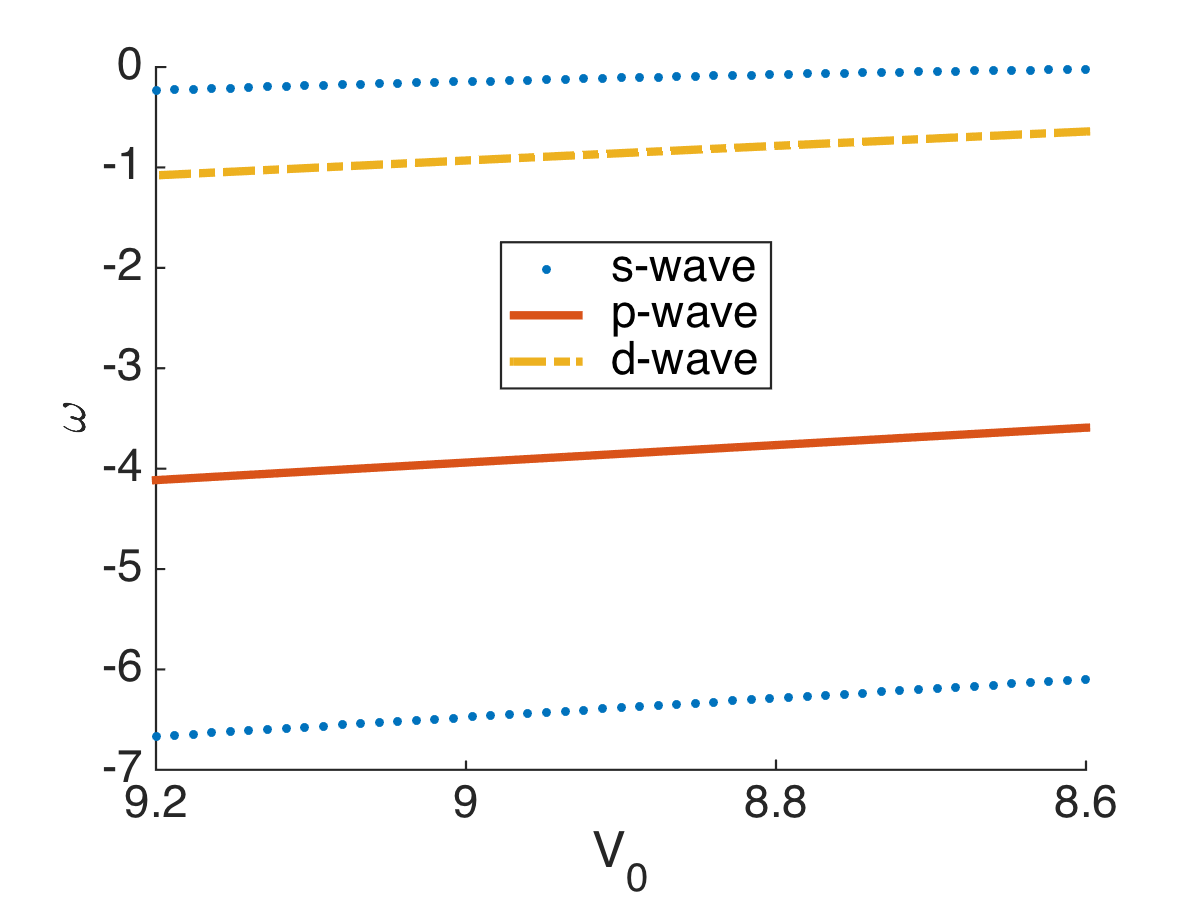}
\caption
{The spectrum of bound states in the square well potential [\eq{squarewell}] with nondimensional width $d=1.15$, as a function of the depth $V_0$. There are two s-waves, one p-wave and one d-wave. The least-bound (s-wave) state reaches the threshold at $V_0 \approx 8.5$ (not shown). In figures \ref{Fig:ImOmegas-wave} to \ref{Fig:MatchingCoefficients} we study the properties of this bound state when a periodic driving force with frequency $\Omega=2$ is turned on.
\label{Fig:Spectrum}}}
\end{figure}

In Fig.~\ref{Fig:Spectrum} we show the spectrum of the time-independent square-well over a small range of $V_0$ values at $d=1.15$. For these parameters, the  external drive fixed at frequency $\Omega=2$ does not resonate with any of the transition frequencies between the states. The least bound s-wave state which we study in the following is pushed towards the threshold at $V_0 \approx 8.5$.

\begin{figure}[ht]
\center {\includegraphics[width=3.0in]{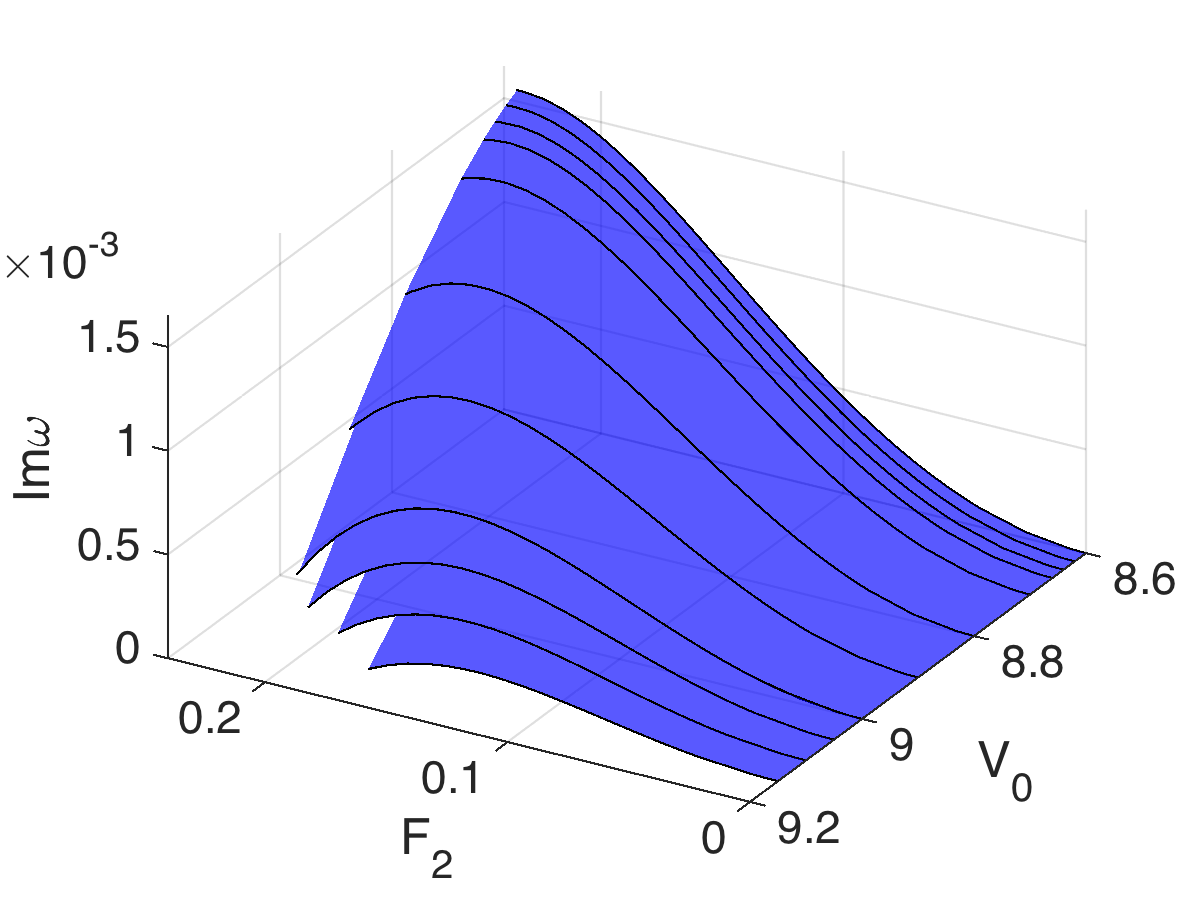}
\caption
{Imaginary part of the Floquet quasi-energy of the state which evolves from the least-bound s-wave state of Fig.~\ref{Fig:Spectrum}, as a function of the periodic drive amplitude and the potential depth. A  nonmonotonous dependence on the parameters can be seen, where beyond some $V_0$-dependent value of $F_2$, an increase of the drive amplitude leads to a decrease of the decay rate of the state out of the well. This decrease terminates when the imaginary part of $\omega$ reaches zero, which happens along a continuous line in parameter space (that here appears ``jittered'' due to the discrete sampling). Beyond this line the resonance seems to have disappeared from the quasi-energy spectrum.
\label{Fig:ImOmegas-wave}}}
\end{figure}

Figure \ref{Fig:ImOmegas-wave} shows the imaginary part of the Floquet quasi-energy of the resonance state which evolves from the least-bound s-wave as a function of $F_2$ and $V_0$. Following the discussion at the end of the previous section, we solve for the state with positive imaginary part, that gives (half) the transition rate for formation of such quasi-bound states. Since the driving force is time-reversal invariant, taking the complex conjugate of this state and reversing the sign of $t$ gives the solution which describes quasi-bound states decaying out of the well. We will therefore (somewhat loosely) refer to the imaginary part of the quasi-energy as the decay rate.

\begin{figure}[ht]
\center {\includegraphics[width=3.0in]{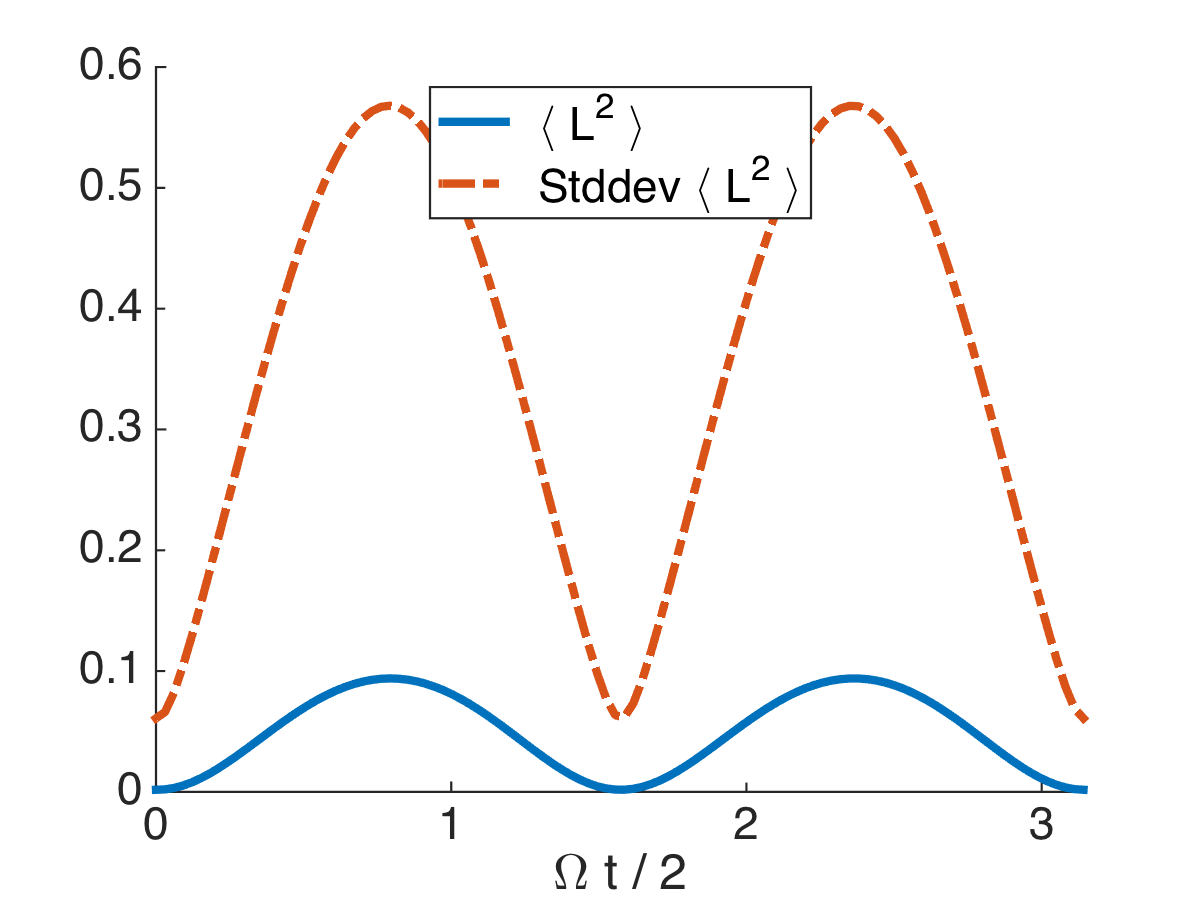}
\caption
{Expectation value and standard deviation of the quantum average of the squared angular momentum operator $\vec{L}^2$ (nondimensional, with $\hbar=1$), of the driven quasi-bound s-wave state. The parameters of the state correspond to the highest point in Fig.~\ref{Fig:ImOmegas-wave}, with $V_0=8.6$ and $F_2=0.235$, and the state's quasi-energy has shifted from $\omega\approx -0.0225$ (see Fig.~\ref{Fig:Spectrum}) to $\omega\approx -0.0612+0.00155i$. The large amplitude oscillations are coherent, while the large standard deviation indicates that the state is an asymptotic broad distribution of different partial waves.
\label{Fig:Lt}}}
\end{figure}

For low drive amplitude, the quasi-bound state's decay rate grows parabolically (as can be inferred from a log-log plot, not shown) which is the expected perturbative result [\eq{epsilon2n}]. In the nonperturbative regime the  decay rate is clearly nonmonotonous; for a strong enough drive the decay rate begins to decrease and then reaches zero, where the two complex-conjugate resonances meet and seem to annihilate and be removed from the quasi-energy spectrum. We note that this is not a threshold effect -- the real part of the quasi-energy is separated from $0$.

In Fig.~\ref{Fig:Lt} we show the  expectation value and the standard deviation of the quantum average of the squared angular momentum operator $\vec{L}^2$ [calculated using eqs.~\eqref{I1integral}-\eqref{L2integral}], for the driven quasi-bound s-wave state at the highest point in the parameter region of Fig.~\ref{Fig:ImOmegas-wave}. Both quantities display a large amplitude oscillation over one period of the drive -- we note that this oscillation in itself is coherent and involves no uncertainty. On the other hand, the probability distribution of the angular momentum (at any fixed time within the period) is seen to be very broad. We can infer that the fact that the expectation value remains close to zero is misleading ($l$ is not a good quantum number even approximately), and under the effect of the drive the state develops a strong superposition of many partial waves. We note that the imaginary part of the energy (which gives a decaying exponential envelope) is ignored here. 

\begin{figure}[ht]
\center {\includegraphics[width=3.0in]{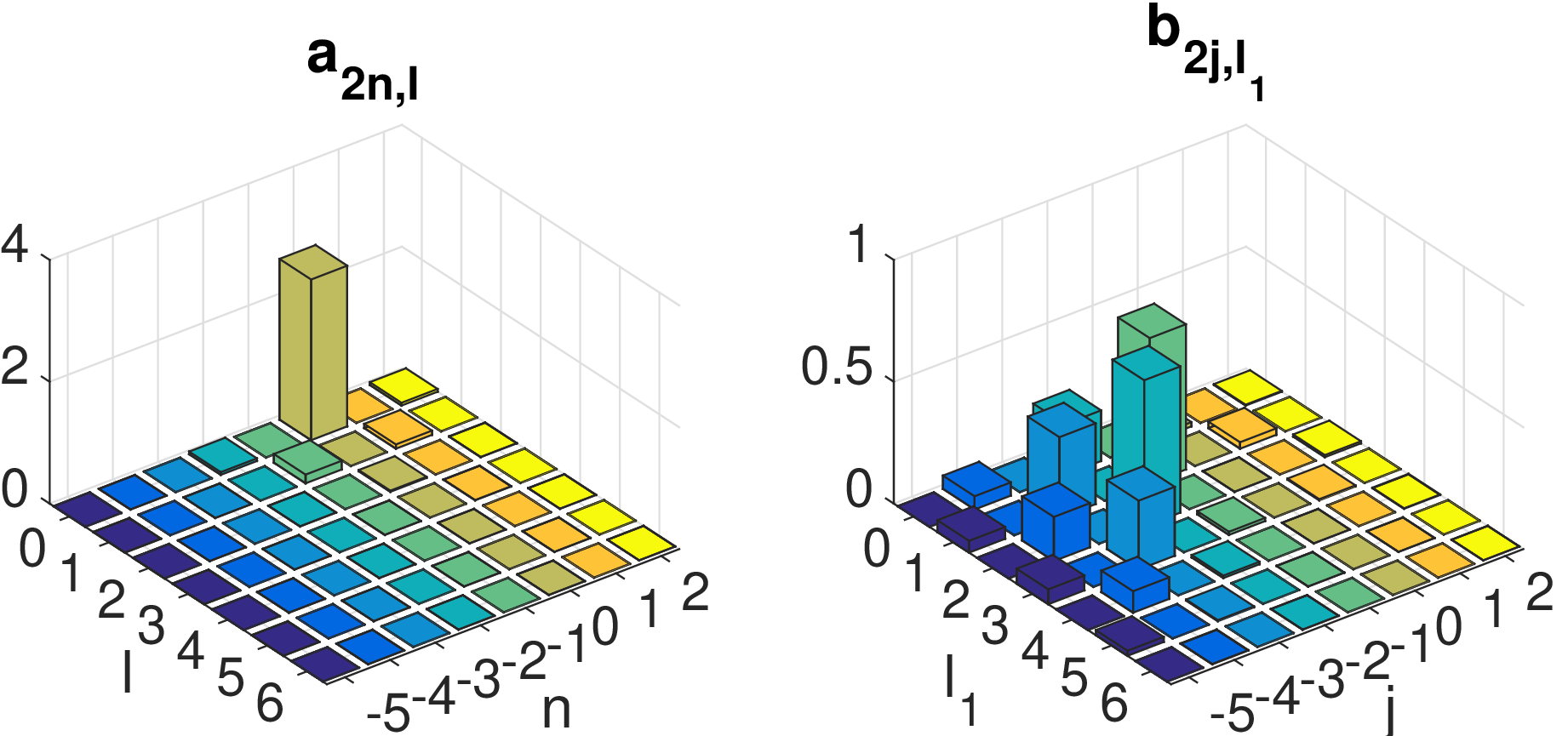}
\caption
{Absolute value of the matching coefficients of the solution ansatz of \eq{phi3Dansatz}, in the interior region ($a_{2n,l}$) and in the exterior region ($b_{2j,l_1}$), for the state of Fig.~\ref{Fig:Lt}. Outside of the range of $\left(n, l\right)$ and $\left(j, l_1\right)$ values shown the coefficients quickly decay. We note that although inside the well the coefficients correspond exactly to a partial wave expansion, in the exterior region the parameter $l_1$ does not correspond directly to a single partial $l_1$-wave for $F_2\neq 0$.
\label{Fig:MatchingCoefficients}}}
\end{figure}

\begin{figure}[ht]
\center {\includegraphics[width=3.0in]{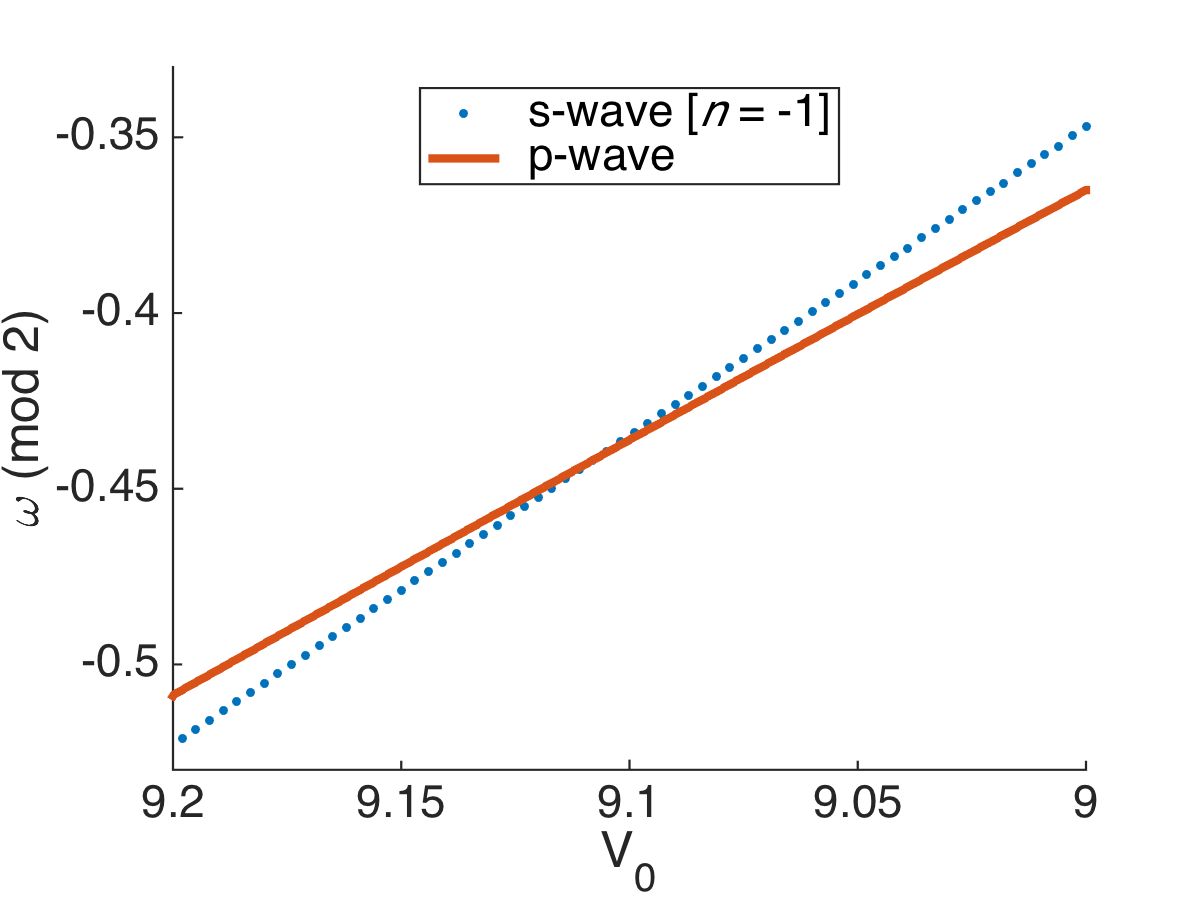}
\caption
{The Floquet quasi-energy $\omega\left({\rm mod\,} 2\right)$ of two bound states in the square potential [\eq{squarewell}] with nondimensional width $d=2.3$ as a function of the depth $V_0$. The s-wave lies in fact deeper in the well than the p-wave for the entire parameter range -- its real energy is $\omega-2$, and the ``crossing'' of the states (calculated for $F_2=0$), becomes relevant only when considering $F_2\neq 0$, when the periodic drive resonates with the energy difference of the two  states, as treated starting with Fig.~\ref{Fig:ReOmega}.
\label{Fig:Spectrum2}}}
\end{figure}

The nature of this superposition can be further seen in the  solution coefficients of the expansion in \eq{phi3Dansatz}, which are depicted in Fig.~\ref{Fig:MatchingCoefficients} for the same state. The quasi-bound s-wave state which for $F_2=0$ would have its entire amplitude at $\left(n=0, l=0\right)$ and $\left(j=0, l_1=0\right)$, has developed a broad superposition of partial waves (mostly outside of the well). The ``checkboard'' pattern is the result of the dipolar nature of the coupling, which conserves $\left(-1\right)^{n+l}$ [or $\left(-1\right)^{j+l_1}$].

\begin{figure}[ht]
\center {\includegraphics[width=3.0in]{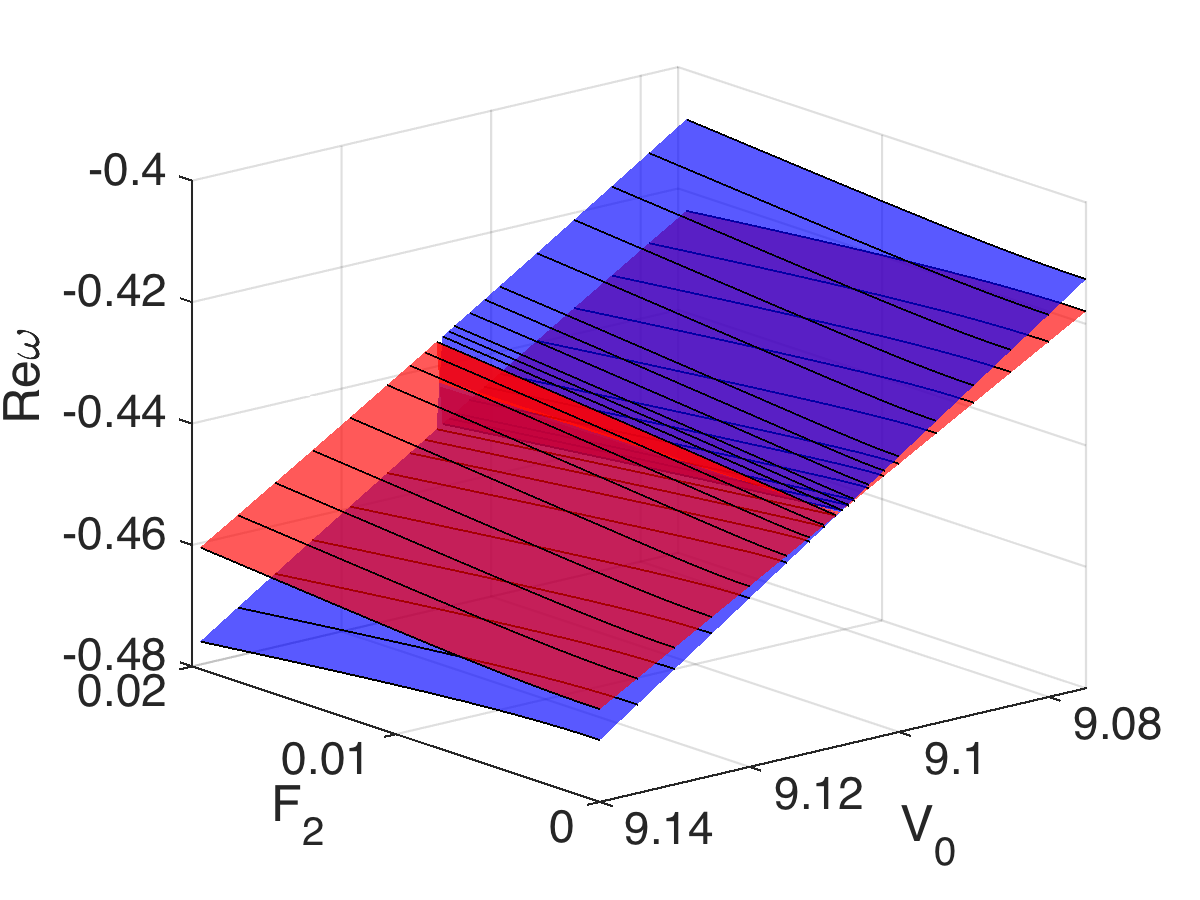}
\caption
{Real part of the Floquet quasi-energy $\omega\left({\rm mod\,} 2\right)$ of the two states of Fig.~\ref{Fig:Spectrum2} in dependence on the depth of the well and the strength of the drive, around the crossing of quasi-energies. Emanating from the crossing point is a singular line of resonance (a ``seem'' of the two surfaces), on which the periodic force mixes completely the two states. The blue surface (lower on the left of the seem, upper on the right of it) corresponds to the state evolving from the s-wave bound state, and the red surface is the p-wave.
\label{Fig:ReOmega}}}
\end{figure}

\begin{figure}[ht]
\center {\includegraphics[width=3.0in]{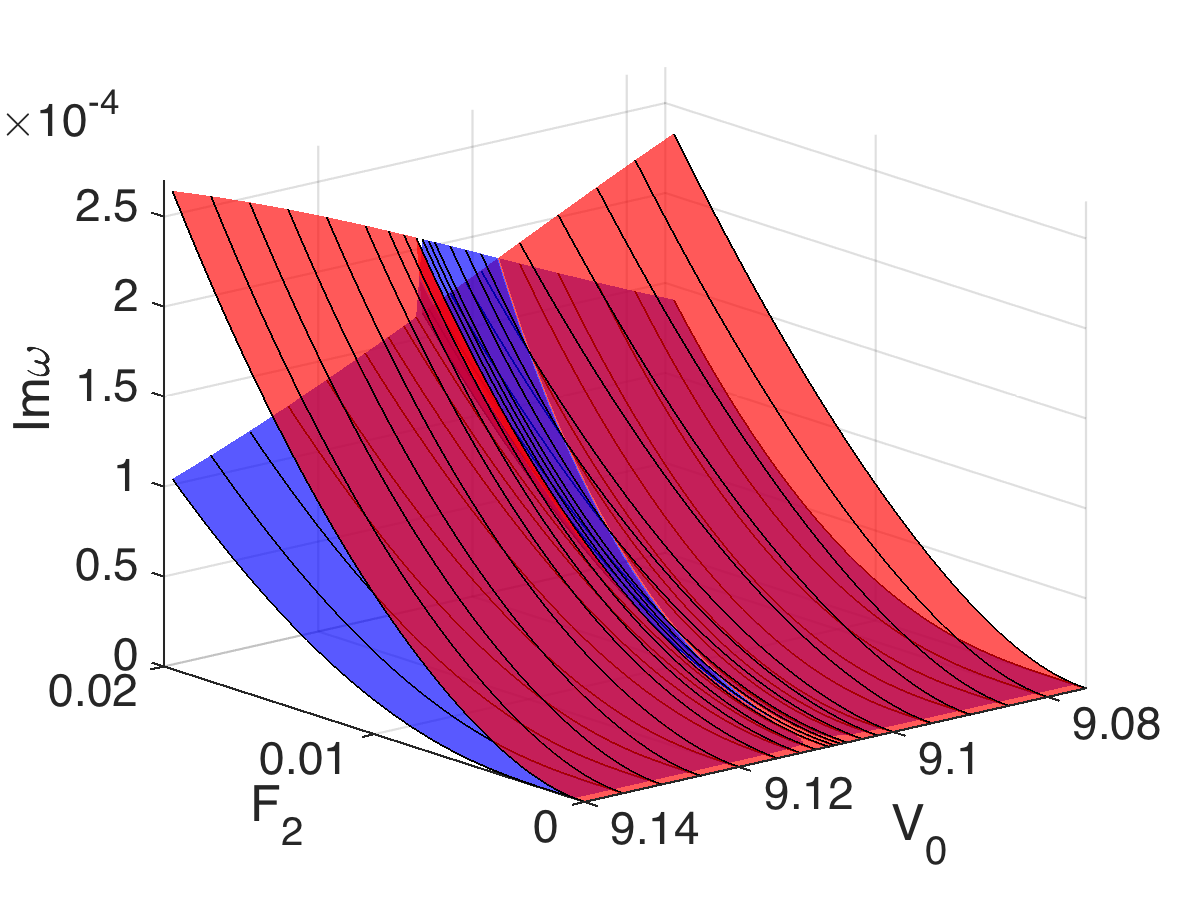}
\caption
{As in Fig.~\ref{Fig:ReOmega}, only showing the imaginary part of the quasi-energy. The s-wave surface is blue, and remains below the p-wave surface for most of the parameter region except in a small part to the right of the seem.
\label{Fig:ImOmega2}}}
\end{figure}

\begin{figure}[ht]
\center {\includegraphics[width=3.0in]{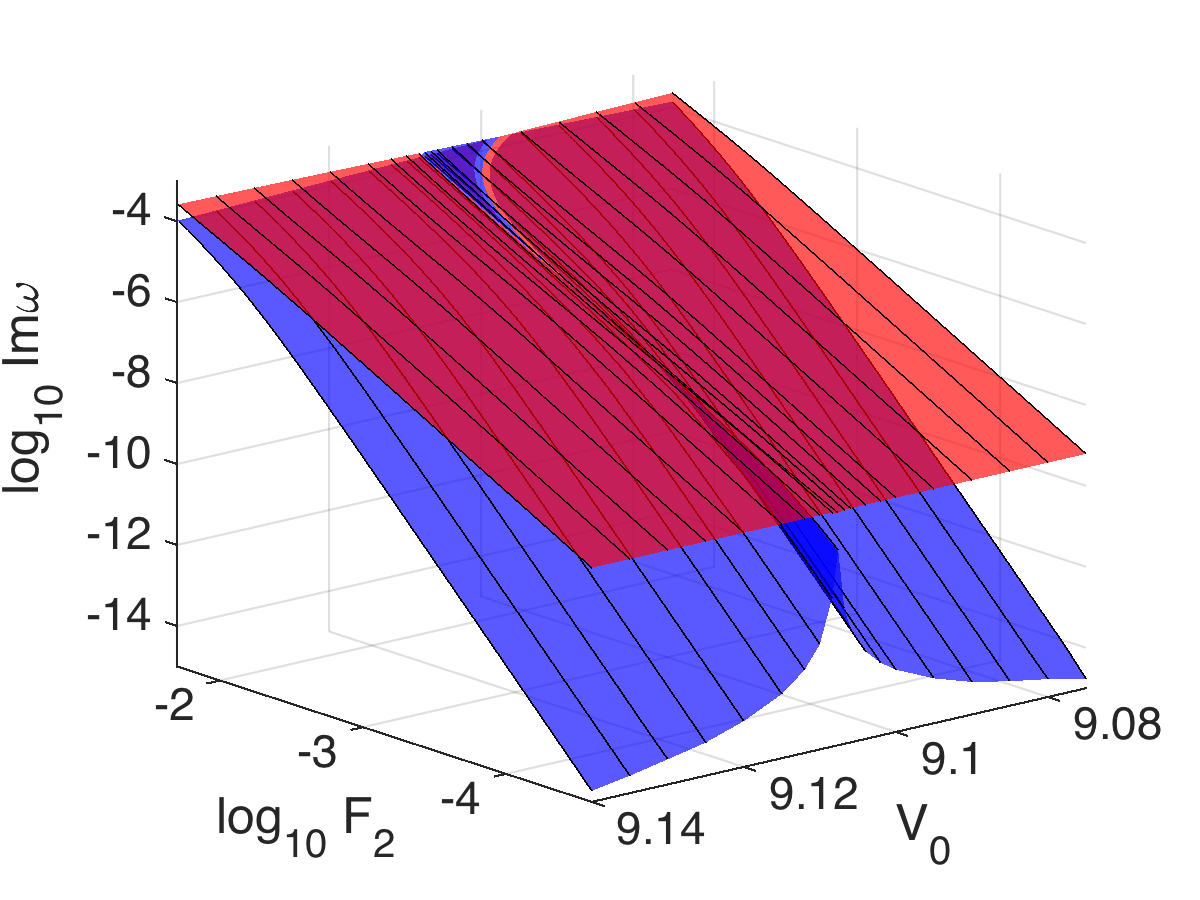}
\caption
{As in Fig.~\ref{Fig:ImOmega2}, only showing the imaginary part of the quasi-energy on a log-log plot. The red surface of the p-wave is now upper for most of the parameter region, with an almost constant slope equal to 2, the perturbation theory result of quadratic decay rate out of the well for this state. The deeper bound s-wave has slope of 4 except within a $V_0$-dependent distance from the resonance, where its strong mixing with the p-wave lowers its power-low exponent to 2.
\label{Fig:ImOmegaLogLog}}}
\end{figure}

Starting with Fig.~\ref{Fig:Spectrum2} we consider a well which is twice wider and supports more bound states. The quasi-energies $\omega\left({\rm mod\,} 2\right)$ of two of the bound states are plotted in this figure as function of $V_0$, around a point of crossing. The energy of the s-wave state is lower by $-2$, so that the crossing is only of quasi-energies (${\rm mod\,} 2$), and is irrelevant for the time-independent well. Figure \ref{Fig:ReOmega} shows the real part of the quasi-energies for the same states, in dependence on both $V_0$ and $F_2$ in a small region of parameters. A resonance line emanates from the crossing point at $F_2=0$, on which the periodic drive's frequency resonates with the energy difference of two states. This line constitutes a singular ``seem'' of the two quasi-energy surfaces (we note that the surfaces can be trivially continued into the $F_2<0$ region).

\begin{figure}[ht]
\center {\includegraphics[width=3.0in]{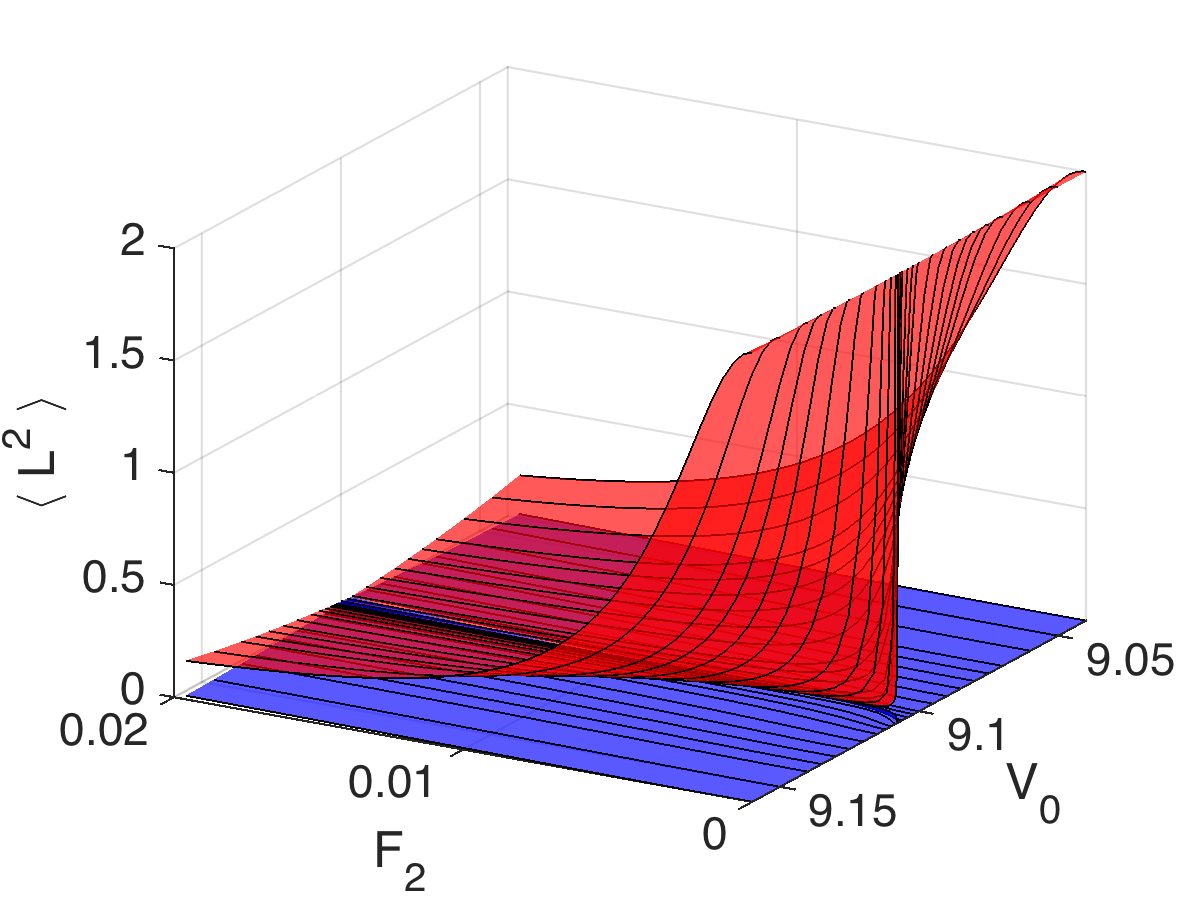}
\caption
{The quantum expectation value of the squared-angular momentum operator for the same states as in figs.~\ref{Fig:Spectrum2}-\ref{Fig:ImOmegaLogLog}, with the range of $V_0$ somewhat larger. The value of $\langle \vec{L}^2 \rangle$ is obtained by averaging over a period of the driving force. The red surface of the p-wave starts at $l\left(l+1\right)=2$ for $F_2=0$ and quickly decreases as a function of $F_2$, depending on the distance to resonance.
\label{Fig:Lmean}}}
\end{figure}

\begin{figure}[ht]
\center {\includegraphics[width=3.0in]{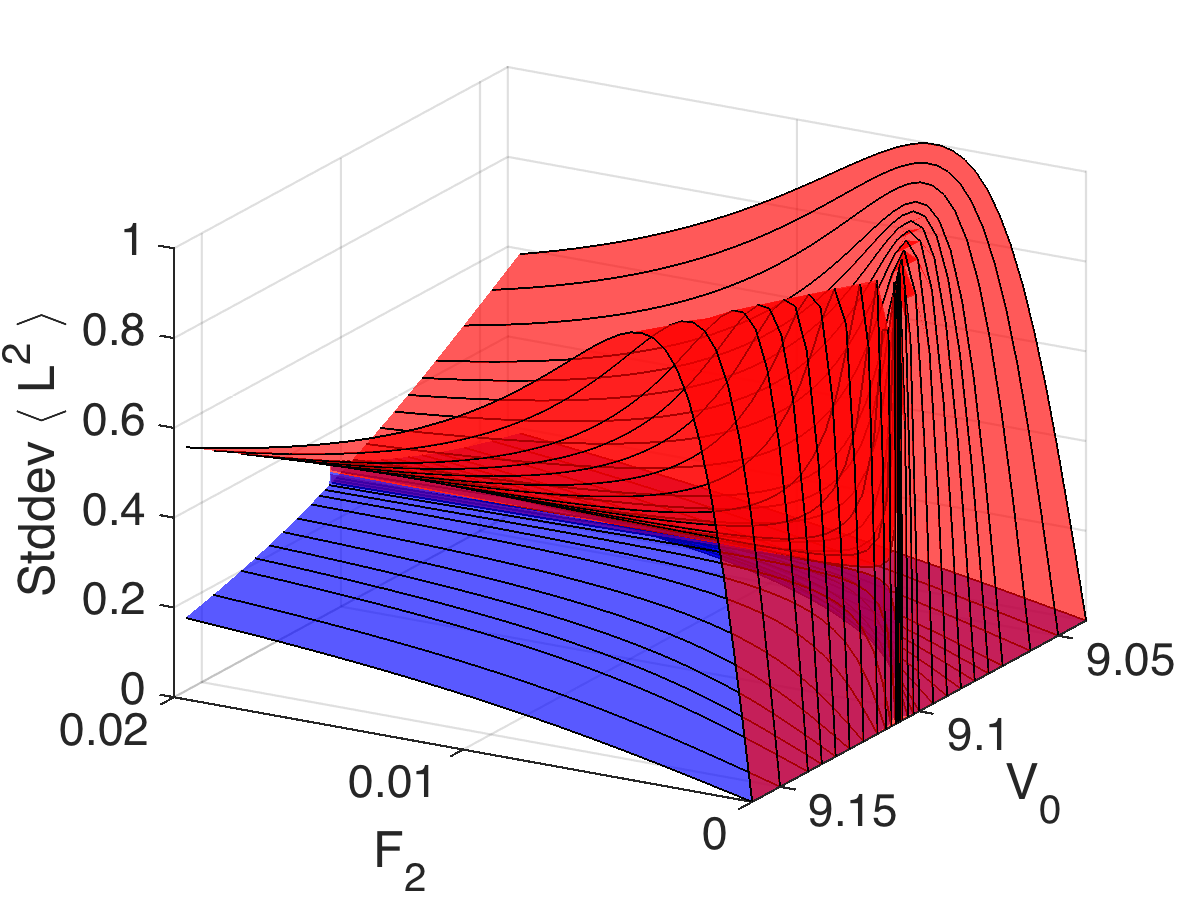}
\caption
{As in Fig.~\ref{Fig:Lmean}, but showing the standard deviation of the squared-angular momentum operator (period-averaged). The p-wave (red) surface grows steeply to its maximal value of 1, achieved after full mixing with the s-wave state, along two lines which emerge from the crossing point in both directions. Beyond these maxima lines, the p-wave is pushed further into an s-wave inside the well, while developing an asymptotic superposition of different partial waves. These two lines can also be identified in Fig.~\ref{Fig:LmeanStdP2P}, and can be considered to border the nonperturbative regime.
\label{Fig:LmeanStd}}}
\end{figure}

\begin{figure}[ht]
\center {\includegraphics[width=3.0in]{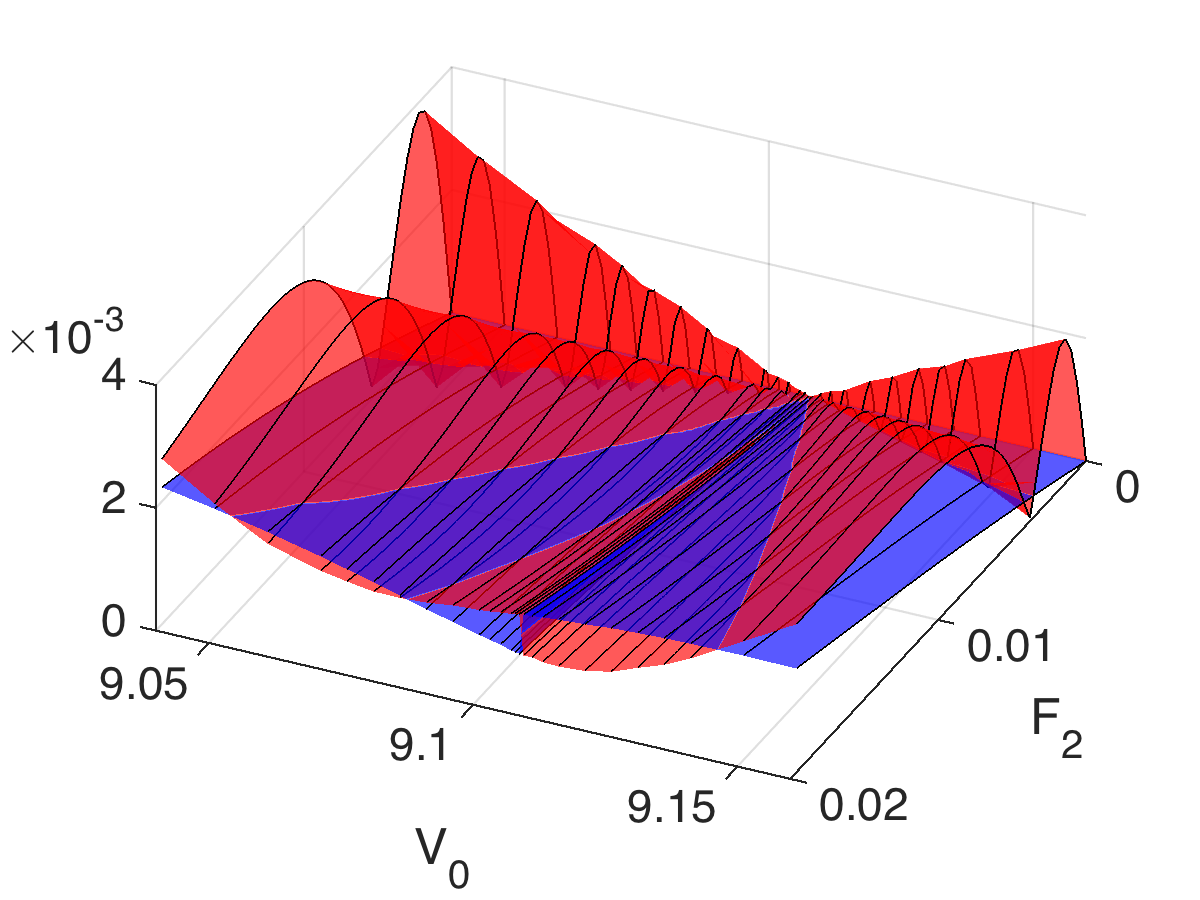}
\caption
{As in Fig.~\ref{Fig:LmeanStd}, but showing the peak-to-peak amplitude of the periodic oscillations of ${\rm Stddev\,}\langle \vec{L}^2 \rangle$, and the axes have been rotated for clarity. The two lines of null oscillation amplitude of the standard deviation correspond to the maxima lines in Fig.~\ref{Fig:LmeanStd} and define the nonperturbative regime.
\label{Fig:LmeanStdP2P}}}
\end{figure}

Figure \ref{Fig:ImOmega2} shows the imaginary part of the quasi-energies of those two states, which is also plotted on a log-log scale in Fig.~\ref{Fig:ImOmegaLogLog}. On the line of degeneracy the two states mix completely. Away from this line, the decay rate of the s-wave, whose real part of the energy lies in the range $-4<\mathfrak{Re}\omega<-2$, grows quartically with $F_2$, while the p-wave's decay rate grows quadratically (and is also much larger), which is again the perturbation theory result. However, close enough to the resonance, it can be seen that the strong partial mixing of the s-wave with the p-wave changes the dependence of the former on $F_2$ to quadratic. The threshold of this regime passes already at $F_2\lesssim 0.01$ for values at the edges of the figures, and  goes down (towards $F_2\to 0$) closer to the resonance. 

Figures \ref{Fig:Lmean} and \ref{Fig:LmeanStd} show the expectation value of $\vec{L^2}$ and its standard deviation, for the same s- and p-wave states over a somewhat larger range of $V_0$ values around the resonance. For $F_2=0$ the value of the nondimensional angular momentum for the s-wave is 0, and for the p-wave it is $l\left(l+1\right)=2$. The standard deviation of the quantum distribution is 0 for both.

The angular momentum expectation value of the p-wave is seen to quickly go down as it is mixed with the s-wave for $F_2\neq 0$, and close enough to the resonance, this happens for arbitrarily small $F_2$. At the same time the standard deviation peaks at 1 for the p-wave (at some $V_0$-dependent value), the result of it being mixed with the s-wave and strongly pushed into the well. Then, for further increase of $F_2$ the p-wave turns more and more into s-wave within the well, while developing a superposition of partial waves outside the well.

\begin{figure}[ht]
\center {\includegraphics[width=3.0in]{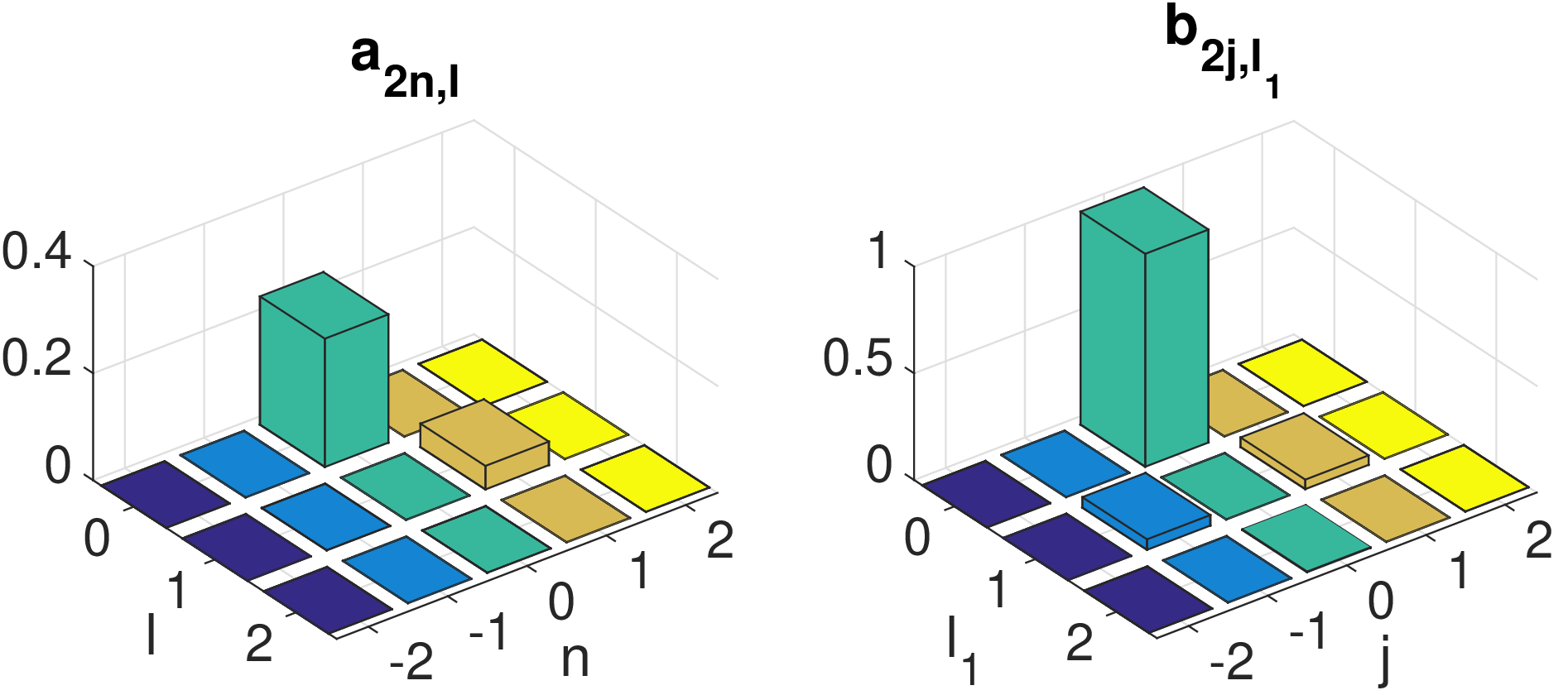}
\caption
{Absolute value of the matching coefficients of the solution (as in Fig.~\ref{Fig:MatchingCoefficients}), for the state continued from the p-wave of figs.~\ref{Fig:Spectrum2}-\ref{Fig:LmeanStdP2P} at $V_0=9.1076$ and $F_2=0.0003$. We note that for $F_2=0$ this state has its entire amplitudes concentrated at $\left(n=1, l=1\right)$ and $\left(j=1, l_1=1\right)$. Already at such a weak drive (but very close to resonance), we find  $\langle \vec{L}^2 \rangle\approx 0.0004$ and ${\rm Stddev\,}\langle \vec{L}^2 \rangle\approx 0.36$, as the state is just entering the nonperturbative regime with amplitudes of new partial waves becoming important in the solution, in addition to the two resonating bound states. With further increase of $F_2$, the state will develop a notable admixture of partial waves, mostly outside of the well.
\label{Fig:MatchingCoefficients2}}}
\end{figure}

The amplitude of the periodic oscillations in the standard deviation of $\vec{L}^2$ is shown in Fig.~\ref{Fig:LmeanStdP2P}. With the driving force being relatively weak ($F_2\lesssim 0.02$), these oscillations (in time) have a small peak-to-peak amplitude (compare with Fig.~\ref{Fig:Lt}). However, the peak-to-peak amplitude itself shows a nontrivial oscillatory pattern in its dependence on the parameters, which is the result of interference of partial waves with different values of $j$ and therefore different factors of $e^{-i2jt}$.

The null lines of the peak-to-peak amplitude, which emanate from the crossing point in both directions in the parameter space and coincide with the maxima lines of the mean standard deviation, can be taken as delimiting the nonperturbative parameter region. In Fig.~\ref{Fig:MatchingCoefficients2} we show the matching coefficients for a state which is very close to resonance and at a very weak drive starts to enter the nonperturbative region. The amplitudes of new partial waves in addition to the two resonating bound states of the time-independent well begin to contribute to the solution, as the drive is strong enough to localize the higher energy p-wave inside the well and quickly dress it with an asymptotic tail of more partial waves.


\begin{acknowledgments}
HL acknowledges support by the French government via the 2013-2014 Chateaubriand fellowship of the French embassy in Israel, support by a Marie Curie Intra European Fellowship within the 7th European Community Framework Programme, and support from COST Action MP1001 (Ion Traps for Tomorrow's Applications), through a Short Term Scientific Mission grant. HL thanks G. V. Shlyapnikov, D. Petrov and P. Rodriguez for fruitful discussions.
\end{acknowledgments}

\begin{widetext}
\appendix

\section{Expansion of linearly-driven cylindrical waves in spherical waves}\label{Sec:Derivations1}

The proof of \eq{intphiRexpansion} proceeds by using \eq{phiRm12} to write
\be
\begin{array}{l} {e^{i\dot{F}^{\pi } \left(t\right)z} \int_{C_1} d\alpha \sin \alpha \chi _{m}^{\left(1\right)} \left(\vec{r};k_{2j} ,\alpha \right)b_{{2j} }^{\left(1\right)} \left(\alpha \right)e^{-iF^{\pi } \left(t\right)k_{2j} \cos \alpha }  }
 \\\\\qquad {=e^{i\dot{F}^{\pi } \left(t\right)z} \int_{C_1} d\alpha \sin \alpha \chi _{m}^{\left(1\right)} \left(\vec{r};k_{2j} ,\alpha \right)\sum _{l_{1} }b_{2j,l_{1} } P_{l_{1} } \left(\cos \alpha \right)\sum _{l_{2} }\left(-i\right)^{l_{2} } \left(2l_{2} +1\right)j_{l_{2} } \left(-F^{\pi } \left(t\right)k_{2j} \right)P_{l_{2} } \left(\cos \alpha \right)   } 
\\\\\qquad {=e^{i\dot{F}^{\pi } \left(t\right)z} \sum _{l_{1} ,l_{2} }b_{2j,l_{1}  } \left(-i\right)^{l_{2} } \left(2l_{2} +1\right)j_{l_{2} } \left(F^{\pi } \left(t\right)k_{2j} \right) \sum _{l_{3} }W\left(P_{l_{1} } ,P_{l_{2} } ,P_{l_{3} }^{m} \right)\int_{C_1} d\alpha \sin \alpha \chi _{m}^{\left(1\right)} \left(\vec{r};k_{2j} ,\alpha \right)P_{l_{3} }^{m} \left(\cos \alpha \right)  } 
\\\\\qquad {=e^{i\dot{F}^{\pi } \left(t\right)z} \sum _{l_{1} ,l_{2} }b_{2j,l_{1}} j_{l_{2} } \left(F^{\pi } \left(t\right)k_{2j} \right)\sum _{l_{3} }c_{l_{1} ,l_{2} ,l_{3} } h_{l_{3} }^{\left(1\right)} \left(k_{2j} r\right)P_{l_{3} }^{m} \left(\cos \theta \right)e^{im\varphi }   } 
\\\\\qquad {=\sum _{l_{4} }i^{l_{4} } \left(2l_{4} +1\right)j_{l_{4} } \left(\dot{F}^{\pi } \left(t\right)r\right)P_{l_{4} } \left(\cos \theta \right)\sum _{l_{1} ,l_{2} }b_{2j,l_{1}} j_{l_{2} } \left(F^{\pi } \left(t\right)k_{2j} \right)\sum _{l_{3} }c_{l_{1} ,l_{2} ,l_{3} } h_{l_{3} }^{\left(1\right)} \left(k_{2j} r\right)P_{l_{3} }^{m} \left(\cos \theta \right)e^{im\varphi }    } 
\\\\\qquad {=\sum _{l_{1} ,l_{2} ,l_{3} ,l_{4} }b_{2j,l_{1}} j_{l_{2} } \left(F^{\pi } \left(t\right)k_{2j} \right)j_{l_{4} } \left(\dot{F}^{\pi } \left(t\right)r\right)h_{l_{3} }^{\left(1\right)} \left(k_{2j} r\right)\sum _{l}c_{l_{1} ,l_{2} ,l_{3} ,l_{4} ,l} Y_{l}^{m} \left(\theta ,\varphi \right)  }
 \end{array}\label{AppEq1}\ee
where the multiplicative factors $e^{-i\frac{1}{2} k_{2j} ^{2} t}$ and $-S_{R,2j,l_1}^{\left(1\right)}$ have been omitted for simplicity, and by using the definition of $R_{2j,l_{1} ,l}^{\left(a\right)} \left(r,t\right)$ given in \eq{R2jl1la}, \eq{AppEq1} results in \eq{intphiRexpansion}. In the derivation of \eq{AppEq1}, the plane-wave expansion in terms of spherical Bessel functions has been used (twice), the coefficients of expansion of a product of two (associated) Legendre polynomials (which can be written using Wigner 3-j symbols) are defined by 
\be W\left(P_{l_{1} }^{m_{1} } ,P_{l_{2} }^{m_{2} } ,P_{l_{3} }^{m_{3} } \right)=\int _{-1}^{1}P_{l_{1} }^{m_{1} } \left(w\right)P_{l_{2} }^{m_{2} } \left(w\right)P_{l_{3} }^{m_{3} } \left(w\right)dw,\label{Wdef}\ee
the coefficients $c_{l_{1} ,l_{2} ,l_{3} } $ are obtained using \eq{hlexpansion} and \eq{Wdef} and given by
\be c_{l_{1} ,l_{2} ,l_{3} } =2\left(2l_{2} +1\right)\left(-1\right)^{l_{2} }i^{l_{2} +l_{3} -1} W\left(P_{l_{1} } ,P_{l_{2} } ,P_{l_{3} }^{m} \right),\ee
and the coefficients $c_{l_{1} ,l_{2} ,l_{3} ,l_{4} ,l} $ are similarly
\be c_{l_{1} ,l_{2} ,l_{3} ,l_{4} ,l} =c_{l_{1} ,l_{2} ,l_{3} } \left(2l_{4} +1\right)i^{l_{4} } W\left(P_{l_{3} }^{m} ,P_{l_{4} } ,P_{l}^{m} \right)/N_{l}^{m},\label{cl1l2l3l4l}\ee
with the definitions
\be Y_{l}^{m} \left(\theta ,\varphi \right)=N_{l}^{m} P_{l}^{m} \left(\cos \theta \right)e^{im\varphi } ,\qquad \qquad N_{l}^{m} =(-1)^m\sqrt{\left(2l+1\right)/4\pi }\sqrt{\left(l-m\right)!/\left(l+m\right)!}.\ee

\section{The projection of two eigenfunctions of the internal Hamiltonian}\label{Sec:Derivations2}

In order to derive \eq{phi1phi2int}, let $\varepsilon _{1} ,\varepsilon _{2} $ be the (possibly complex) energies of two complex eigenfunctions $\phi _{1} ,\phi _{2} $ of the  interior Hamiltonian $H_{L}$. For the projection of the two within the interior region, we can write (since both sides vanish)
\be 0=\left\langle \phi _{2} \left|\left(H_{L} -\varepsilon _{1} \right)\left|\phi _{1} \right. \right. \right\rangle -\left\langle \phi _{1} \left|\left(H_{L} -\varepsilon _{2} \right)\left|\phi _{2} \right. \right. \right\rangle.\ee
By canceling the potential energy terms, we get after rearranging the kinetic terms and terminating the integration at an arbitrary point $d$ (which is allowed since the equality above holds identically in space),
\be \begin{array}{l} {\int _{0}^{d }\left(\varepsilon _{1} \phi _{2} ^{*} \phi _{1} -\varepsilon _{2} \phi _{1} ^{*} \phi _{2} \right)r ^{2} dr  =-\frac{1}{2} \int _{0}^{d}\left(u_{2} ^{*} \partial _{r } ^{2} u_{1} -u_{1} ^{*} \partial _{r } ^{2} u_{2} \right)dr  } \\\\ {\qquad =-\frac{1}{2}  \left[\left. \left(u_{2} ^{*} \partial _{r } u_{1} -u_{1} ^{*} \partial _{r} u_{2} \right)\right|_{d } -\int _{0}^{d }\left(\partial_{ r } u_{2} ^{*} \partial _{r} u_{1} -\partial _{r } u_{1} ^{*} \partial _{r } u_{2} \right)dr  \right]} \end{array}\ee
where the factor of $1/2$ appears since we assume that the nondimensional kinetic energy term is $-\frac{1}{2}\nabla^2$. In the second line of the above equation, the integrated term is purely imaginary being the difference of two complex conjugates. Taking the complex conjugate of the entire equation and adding, this term drops and we get
\be \int _{0}^{d }2\,\mathfrak{Re}\left\{\left(\varepsilon _{1} -\varepsilon _{2} \right)\phi _{2} \phi _{1} ^{*} \right\}r ^{2} dr  =-\frac{1}{2}  2\,\mathfrak{Re}\left. \left\{u_{2} \partial _{r } u_{1} ^{*} -u_{1} \partial _{r } u_{2} ^{*} \right\}\right|_{d},\ee
which gives immediately \eq{phi1phi2int}.

\section{The expectation value of tensor operators}\label{Sec:Derivations3}

In this appendix we give explicitly the expansion of integrals which are required in order to calculate expectation values of general tensor operators, in the Floquet eigensolutions of \seq{Sec:Matching3D}. For simplicity we treat  here only the most useful case of axially symmetric wavefunctions, with $m=0$ (no $\varphi$ dependence). 

Using the notation introduced in \eq{phi1ruY}, we start by writing the $\pi$-periodic part of the wavefunction in the form
\be \phi ^{\pi } \left(\vec{r} ,t\right) = \sum_{n,l}a_{2n,l}e^{-i2nt}\frac{1}{r}u ^{\pi }_{2n,l} \left(r,t\right) Y_l^0,\label{phipiint} \ee
which corresponds to the expansion in \eq{phi3Dansatz} of wavefunctions in the interior region. For such wavefunctions, we define the (unnormalized) expectation value in the interior region of a purely radial operator $\mathcal{O}\left(r\right)$,
\be \mathfrak{I}_0\left[\mathcal{O}\left(r\right)\right]\equiv\int d^3\vec{r}\left|\phi^{\pi} \left(\vec{r}, t\right)\right|^{2} \mathcal{O}\left(r\right) = \sum_{l,l'} \delta_{l,l'} \sum_{n,n'}e^{2i\left(n-n'\right)t}a_{2n,l}^*a_{2n',l'}\int d\xi_{a}\left[u_{2n,l} ^{\pi } \right]^{*}  \mathcal{O}\left(r\right) u_{2n',l'} ^{\pi }. \ee
The above expression can be rewritten as
\be \mathfrak{I}_0\left[\mathcal{O}\left(r\right)\right]= \sum_{l}   I_{l,l}\left[\mathcal{O}\left(r\right)\right],\ee
where we have defined for convenience the functional
\be {I}_{l,l'}\left[\mathcal{O}\left(r\right)\right]=   \sum_{n\leq n'}\left(2-\delta_{l,l'}\delta_{n,n'}\right)\,\mathfrak{Re}\left\{e^{2i\left(n-n'\right)t}a_{2n,l}^*a_{2n',l'}\int dr\left[u_{2n,l} ^{\pi } \right]^{*}  \mathcal{O}\left(r\right) u_{2n',l'} ^{\pi }\right\},\label{Ill}\ee
with the summation taken over pairs of states enumerated by $\left\{\left(n,l\right),\left(n',l'\right)\right\}$ with fixed $l$ and $l'$.

For example, the normalization integral calculated for any time (generalizing \eqref{normalization3D} for expansions with free-particle components) can be written using the above notation as
\be \mathfrak{I}_0\left[\hat{1}\right]=\sum_{l} I_{l,l}\left[\hat{1}\right],\label{I1integral}\ee
with $\hat{1}$ the identity operator. Any other expectation value must then be divided by the value of this normalization integral. Similarly, the expectation value of the squared angular momentum operator $\vec{L}^2$ is given by
\be \mathfrak{I}_0\left[\vec{L}^2\right]=\sum_{l} l\left(l+1\right) I_{l,l}\left[\hat{1}\right].\label{L2integral}\ee

For an operator of a general radial part multiplied by the position vector, $\mathcal{O}\left(r\right)\vec{r}$, only the Cartesian $z$-component survives the integral (for axially symmetric wavefunctions), and we can write using $ z/ r=\cos\theta$
\be \mathfrak{I}_1\left[\mathcal{O}\left(r\right)\vec{r}\,\right] = \int d^3\vec{r}\left|\phi^{\pi} \left(\vec{r} ,t\right)\right|^{2} \mathcal{O}\left(r\right)\vec{r} = \sum_{l,l'}p_{l,l'}  \sum_{n,n'}e^{2i\left(n-n'\right)t}a_{2n,l}^*a_{2n',l'}\int dr\left[u_{2n,l} ^{\pi } \right]^{*}  \mathcal{O}\left(r\right)r\, u_{2n',l} ^{\pi },\ee
with the coefficients being
\be p_{l,l'} =2\pi N_{l}^{0}N_{l'}^{0}\int d\theta \sin \theta \cos \theta P_{l} \left(\cos \theta \right)P_{l'} \left(\cos \theta \right).\ee
Using that fact that nonzero terms will have $\left|l-l'\right|=1$, we find
\be \mathfrak{I}_1\left[\mathcal{O}\left(r\right)\vec{r}\,\right] =  \hat{z}\sum_l p_{l,l+1}I_{l,l+1}\left[\mathcal{O}\left(r\right)r\right].\ee

For an operator with a general radial part multiplied by a bilinear combination of position vector components, $\mathcal{O}\left(r\right)\vec{r}_{\alpha}\vec{r}_{\beta}$, where $\alpha,\beta\in\left\{x,y,z\right\}$, only the diagonal terms with $\alpha=\beta$ survive the integration (for wavefunctions with $m=0$), with the result
\be \mathfrak{I}_2\left[\mathcal{O}\left(r\right)\vec{r}_{\alpha}\vec{r}_{\beta}\right] =\delta_{\alpha,\beta} \sum_{l,l'}q_{\alpha,l,l'} I_{l,l'}\left[\mathcal{O}\left(r\right)r^2\right],\ee
where
\be q_{\alpha,l,l'} =2\pi N_{l}^{0}N_{l'}^{0}\int d\theta \sin \theta \left[ \cos^2 \theta\delta_{\alpha,z}+\frac{1}{2} \sin^2 \theta\left(\delta_{\alpha,x}+\delta_{\alpha,y}\right) \right] P_{l} \left(\cos \theta \right)P_{l'} \left(\cos \theta \right).\ee

In all of the above expressions, ${I}_{l,l'}\left[\mathcal{O}\left(r\right)\right]$ as defined  in \eq{Ill} is valid in the interior region. To get the complete result for expectation values in whole space, the integration over the exterior region must be added, where the wavefunctions are expanded differently in \eq{phi3Dansatz}. In this case, \eq{phipiint} is to be replaced by 
\be \phi ^{\pi } \left(\vec{r} ,t\right) = \sum _{ j,l_{1}} b_{2j,l_{1} }e^{-i2jt}\sum _{l}\frac{1}{r}u_{2j,l_1,l}^{\pi} \left(r,t\right)Y_{l}^0\label{phipiext},\ee
and accordingly, \eq{Ill} becomes in the exterior region
\be {I}_{l,l'}\left[\mathcal{O}\left(r\right)\right]= \sum_{j,j',l_{1} ,l_{1} ' }e^{2i\left(j-j'\right)t} b_{2j,l_{1} }^*b_{2j',l_{1} ' }
\int dr\left[u_{2j,l_1,l} ^{\pi } \right]^{*}  \mathcal{O}\left(r\right) u_{2j',l_1',l'} ^{\pi }.\label{Illext}\ee

\end{widetext}

\bibliographystyle{../h-physrev}

\bibliography{../scattering}

\end{document}